\documentclass[conference]{IEEEtran}
\IEEEoverridecommandlockouts

\usepackage{tikz}
\usepackage{amsmath}

\usepackage{comment}
\usepackage{endnotes}

\usepackage{footnote}

\usepackage[normalem]{ulem}
\usepackage{url}
\usepackage[T1]{fontenc}
\usepackage{float}
\usepackage[T1]{fontenc}
\usepackage{multirow}
\usepackage{tabu}
\usepackage{xspace}
\usepackage{todonotes}

\newcommand{\projname}{{\sc Extractor}\xspace}
\usepackage{adjustbox}
\usepackage{lipsum}
\usepackage[htt]{hyphenat} 
\usepackage{array}
\usepackage{pgfplots}
\usepackage{tikzpeople}
\usepackage{graphicx}
\usepackage{tikz}
\usepackage{dblfloatfix}
\usepackage{times}
\usepackage{enumitem}
\usepackage{makecell}
\usepackage[colorlinks=true,urlcolor=black]{hyperref}
\def\BibTeX{{\rm B\kern-.05em{\sc i\kern-.025em b}\kern-.08em
    T\kern-.1667em\lower.7ex\hbox{E}\kern-.125emX}}
    
\newcolumntype{?}{!{\vrule width 1.5pt}}

\def\checkmark{\tikz\fill[scale=0.4](0,.35) -- (.25,0) -- (1,.7) -- (.25,.15) -- cycle;}

\usepackage{hyperref}
\usepackage{cleveref}

\crefformat{section}{\S#2#1#3} 
\crefformat{subsection}{\S#2#1#3}
\crefformat{subsubsection}{\S#2#1#3}

\title{EXTRACTOR: Extracting Attack Behavior from Threat Reports}

 \author{
 {\rm Kiavash Satvat}\\
 University of Illinois at Chicago\\
 ksatva2@uic.edu
 \and
 {\rm Rigel Gjomemo}\\
 University of Illinois at Chicago\\
 rgjome1@uic.edu
 \and
 {\rm V.N. Venkatakrishnan}\\
 University of Illinois at Chicago \\
 venkat@uic.edu
 }

\date{}

\begin{document}

\newcommand{\vtodo}[1]{\todo[inline]{Venkat: #1}}
\newcommand{\rtodo}[1]{\todo[inline]{Rigel: #1}}
\newcommand{\ktodo}[1]{\todo[inline]{Kiavash: #1}}

\maketitle
\begin{abstract}
\noindent The knowledge on attacks contained in Cyber Threat Intelligence (CTI) reports is very important to effectively identify and quickly respond to cyber threats. However, this knowledge is often embedded in large amounts of text, and therefore difficult to use effectively. To address this challenge, we propose a novel approach and tool called \projname that allows precise automatic extraction of concise attack behaviors from CTI reports.  \projname makes no strong assumptions about the text and is capable of extracting attack behaviors as provenance graphs from unstructured text. We evaluate \projname using real-world incident reports from various sources as well as reports of DARPA adversarial engagements that involve several attack campaigns on various OS platforms of Windows, Linux, and FreeBSD.  Our evaluation results show that \projname can extract concise provenance graphs from CTI reports and show that these graphs can successfully be used by cyber-analytics tools in threat-hunting. 

\end{abstract}

\pagestyle{plain}

\section{Introduction}

Cyber Threat Intelligence (CTI), as commonly reported in technical reports, whitepapers, blogs, and newsgroups, is a valuable source of information about cyber-attacks. These reports describe many aspects of an attack in natural language, including the sequence of actions, effects on the system under attack, and  Indicators of Compromise (IOC).  The knowledge contained in CTI reports is essential for cyber operations and response personnel, system administrators, as well as vendors of intrusion detection and prevention products.  

Previous studies \cite{liao2016acing,husari2017ttpdrill,zhu2018chainsmith} utilize various NLP techniques to automatically extract knowledge available in CTI reports in the form of IOCs (i.e., \cite{liao2016acing,zhu2018chainsmith}) and threat actions (i.e., \cite{husari2017ttpdrill}).
While these works provide a good starting point towards automated extraction of threat elements (IOCs and threat actions) from CTI reports, they do not extract the relationships between IOCs and threat actions, in order to provide a comprehensive view of the attack behavior. Such attack behavior extraction is essential in threat-hunting activities. 
In particular, extracting attack behavior and the attack's big picture requires extracting the entities involved (e.g., files and sockets), actions (e.g., system calls), the causal and temporal dependencies between them, as well as information flow between the entities. Extracting the attack behavior requires an approach that is able to understand {\em “who did what to whom”, “when” and “where”} from the natural text. This task presents several challenges.

{\em Challenge 1: Verbosity.} Threat reports are infused with a significant amount of irrelevant text; often, only a small portion of the report describes attack behavior. For instance, a description of the malware's geographical origin, though interesting, does not contribute to the description of the malware behavior in a system.  

\begin{figure*} []
	\centering
	\vspace{-44mm}
	\includegraphics[width=1\textwidth]{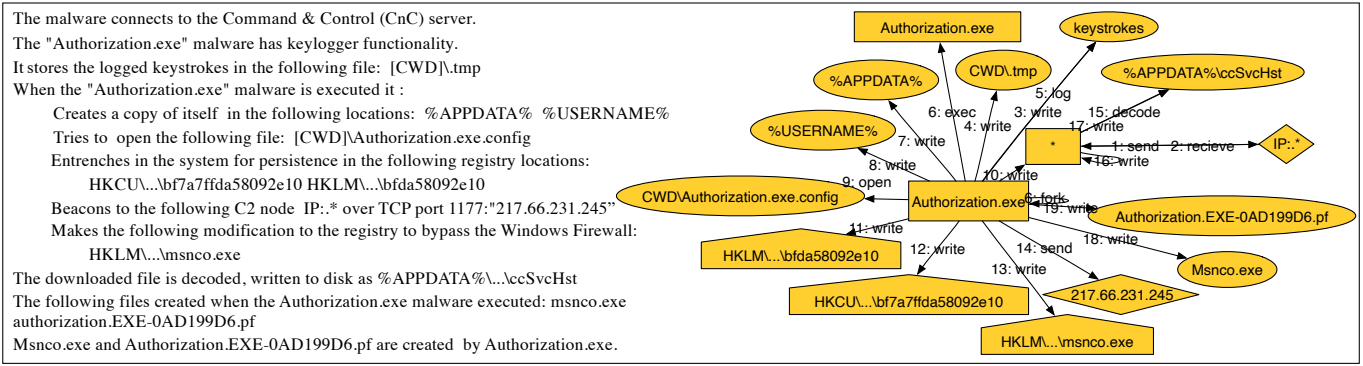}
	\vspace{-50mm}
	\caption{The report (the left) is a free adaptation from the njRAT, where the irrelevant sentences are removed. This example demonstrates the language complexities, which will be discussed throughout the paper. In the corresponding provenance graph (the right), nodes signify system entities, and the edges point to system calls.  The rectangle, oval, pentagon, and diamond represent the file, process, registry, and socket, respectively.}
	\label{fig:background}
\end{figure*}

{\em Challenge 2. CTI text complexity} An important assumption of the previous approaches is that the text structure of CTI reports is (a) relatively simple \cite{liao2016acing}  or (b) that it follows a specific grammatical structure \cite{husari2017ttpdrill} or (c) assuming some patterns in describing concepts \cite{zhu2016featuresmith} or (d) considering stable grammatical relations in the presentation of the sentence in the form of subject, verb and object \cite{liao2016acing,husari2017ttpdrill}.  While these assumptions do not interfere with the goal of these works to extract IOCs and threat action in isolation, in fact, the majority of CTI reports contain much more complex domain-specific contexts (see Section~\ref{sec:background}), which makes the extraction of attack behavior and causal inference more challenging.  
The CTI reports' syntactic and semantic complexities, the prevalence of technical terms, and lack of proper punctuation in these reports \cite{mu2018understanding} can easily impact the interpretation of the report and extraction of attack behavior.

 {\em Challenge 3. Relationship Extraction.}  IOCs and threat actions can be extracted using approaches like string matching and classifiers, as suggested by \cite{liao2016acing} and \cite{husari2017ttpdrill}. However, extracting the big picture, while maintaining concise causal, temporal, and information flow of the attack throughout the report is far more complex and challenging.  In fact, accurately interpreting the complex logic in technical reports is known to be an open problem in NLP \cite{mu2018understanding}.

In this paper, we introduce \projname{\footnote{https://github.com/ksatvat/Extractor}}, which addresses these challenges. The  {\em main goal} of \projname is to concisely extract the full picture of the attack behavior from the technical reports in the form of a graph.  
\projname overcomes the first challenge by proposing a novel text summarization approach that discerns the attack behavior text from the rest. To overcome the second challenge and to optimize overall system performance,  \projname uses a set of techniques to transform a highly complex text into a more consumable form. To address the third challenge, \projname uses a novel approach Semantic Role Labeling (SRL), which allows us to extract the attack behavior and subject, object, and actions of the sentence by inferring the fact of {\em “who did what to whom”, “when” and “where”} (details of these steps discussed in Sections \ref{sec:background} and \ref{sec:methodology}). Finally, the result of SRL in the final step is presented in the form of a {\em graph} describing the attack's steps, artifacts, the causal information flow between the entities involved.


In addition to the main goal of extracting the full attack picture, \projname  follows two more goals:

\noindent

\noindent
{\em Goal 1: Actionable Intelligence}. We want to automatically construct what we call {\em actionable} intelligence. We want to extract from a CTI report only information that is {\em ready} to be used for {\em detection}, or {\em threat hunting}. This means that the attack behaviors we extract from the text must be {\em observable} in the system audit logs and can be effectively used for threat detection. This is an important goal for every approach that extracts attack information from CTI reports. In fact,  we envision the deployment of \projname  as a first step in a {\em threat hunting} operation.

\noindent
{\em Goal 2: Process  a Large Number of CTI Reports accurately}.
We want to process a large number of CTI report, blogs, and attack descriptions from threat detection centers. Accomplishing this goal would enable analysts to automatically tap from a vastly larger source of knowledge than it is currently possible.

\noindent
\textbf{Applications of EXTRACTOR}.
As has been widely demonstrated, the presence of a concise attack behavior description is preferable to have a simple collection of IOCs  in detecting threats~\cite{milajerdi2019poirot,sun2018using,hassan2019nodoze,ma2015accurate,kwon2018mci,ma2016protracer}. 
 \projname is able to build graph representations that represent such concise description from CTI reports, thus guiding intrusion detection and threat hunting systems. Another envisioned use of \projname is that of extracting information from a variety of CTI sources related to the same attack in different organizations. This is to obtain a complete picture of how the same malicious actor might behave in different scenarios.

\projname surpasses the state of the art significantly by making several important contributions. In particular, \projname: 1) significantly expands the range of CTI reports that can be processed, 2) extracts significantly more complex details than the previous studies (e.g., \cite{liao2016acing,husari2017ttpdrill,zhu2018chainsmith}); this includes extraction of causal dependency and temporal order of attack,  3) implements a novel application of extracting semantic relationships among artifacts of an attack that enables it to obtain a much better picture of the attack, 4) implements several novel applications of text simplification and reduction (or summarization)  that enable condensing the text without losing useful information.

This paper is organized as follows. In Section \ref{sec:background}, we provide a more detailed description of the problem and some background information. In Section \ref{sec:methodology}, we describe our approach in detail. In Section \ref{sec:implementation}, we give a short overview of the implementation and different tools used. Section \ref{sec:evaluation} presents the evaluation. Section \ref{sec:discussion} provides a discussion, while Sections \ref{realated} and \ref{conclusion} contain related work and conclusions, respectively.

\section{Problem and Background}\label{sec:background}

\subsection{Problem Description}\vspace{-2mm}

As mentioned in the introduction, the main goal of this paper is to extract {\em actionable} graphs representing attack behavior from generic CTI reports. By {\em actionable} we refer to the important goal of using the extracted knowledge as a signal in {\em threat hunting}. We refer to these graphs as {\em provenance graphs}.  Provenance graphs are a common representation of kernel audit logs \cite{king2005enriching,king2003backtracking}. They represent events (system calls) in a system as edges between entities (processes, files, sockets). Provenance graphs have recently been successfully used for threat detection and  forensic analysis in a large number of studies \cite{hossain2017sleuth,milajerdi2019poirot,sun2018using,hassan2019nodoze,ma2015accurate,kwon2018mci,ma2016protracer}. 

An example of the text contained in CTI reports, inspired by the njRAT attack \cite{solutions2013njrat}, is shown in Figure \ref{fig:background}. This example will be used throughout the paper to illustrate different aspects of our approach. An example of the corresponding provenance graph extracted from that text is also shown in Figure \ref{fig:background} on the right side. As can be noticed, the provenance graph contains nodes that represent entities (processes, registry keys, etc.) involved in the attack and edges that represent the actions carried out by those nodes. In addition, the names of the nodes are such that can be observed in the audit logs, and edges connecting the nodes represent system calls that also appear in the audit logs (goal 1: actionable intelligence). In addition, the graph contains only attack behavior-related nodes and no other information (main goal of full attack picture and conciseness). We note that the natural text in Figure \ref{fig:background} does not have any particular structure (goal 2: process CTI reports written in natural language).

There are several challenges in extracting concise and {\em actionable} provenance graphs from CTI reports written in natural language. First, we need to distinguish attack behavior text from the rest of the report. This implies extracting from the natural text only the kind of relations that describe attack behavior and that can be {\em observed} in audit logs, while filtering out the rest of the text.   Therefore, we need to understand the relations and actions occurring among system entities mentioned in the text to map those actions to system calls, which are represented as edges in the provenance graphs.  Second, we need to overcome CTI text complexities, which may impact our graph extraction. This implies resolving different kinds of ambiguities and complexities present in natural language writing.
We describe the challenges that must be solved in more detail next.   

\subsection{Challenges}\label{subsec:challenges}

\noindent
\textbf{Verbosity.}
In general, CTI reports can be verbose. Sentences containing useful information may be nested inside the text that is not strictly related to the attack, e.g., introductory details. For instance, out of 42 pages DustySky report \cite{DustySky}, only 11 sentences describe the actual attack behavior that can be observed in audit logs. We separate useful content from non-useful using a novel {\em summarization} technique (See \ref{subsec:ats}).

\noindent
\textbf{CTI Text Complexity.} \label{subsec:charac}
The language used in the cybersecurity domain has several peculiarities that NLP tools/techniques (developed for more generic domains) often struggle with. This makes it challenging to use these tools as they are. We list some of these peculiarities below.

\noindent
 {\em Punctuation.} Many CTI reports do not use sentence-ending characters \texttt{`.,!,?'} to delimit sentences. This makes it hard for the popular NLP toolkits, such as Stanford~\cite{stanfordsentence}, NLTK~\cite{loper2002nltk}, and spaCy~\cite{spacy}, to understand the {\em real} sentence boundaries in CTI reports, resulting in texts with long sentences, each of which contains several shorter sentences. For instance, from the observation of 4020 threat reports from the Microsoft threat report center, we notice that writers tend to pack many actions within one sentence, therefore making the average sentence length equal to 52 words (with some examples as long as 313 words per sentence). In contrast, the average English text on which NLP tools are usually trained and designed for, contains approximately 14.4 words per sentence \cite{avrageelen}.

\noindent
{\em Domain-specific words.} Words denoting objects in the cybersecurity domain may have different meanings and contexts from words used in the common English language, on which NLP tools are trained. For instance, IP addresses, paths, process names, system call names, and many other terms often are misunderstood by common NLP tools. This challenge must be met by a mechanism that brings domain insight to assign meanings to the terms. 

\noindent
{\em Ellipsis.} This term denotes a gap in a sentence that: 1) has a missing subject, or 2) has a missing object \cite{chen2016chinese}. This structure is not common in natural English writing \cite{nariyama2004subject}, but it is very common in CTI reports where attacks are described as sequences of actions. For instance, {\tt Creates a copy of itself in the following locations} in Figure \ref{fig:background}  represents an example of {\em ellipsis subject}. 

\noindent
{\em Pronouns.} Pronouns are very commonly used in English \cite{pennebaker2011secret}. Ignoring pronouns may result in their appearance as nodes in the provenance graph in the place of the referent entities.  

\noindent
{\em Other linguistic structures.} Structural complexities and the use of various linguistic techniques such as anaphora, nominalization, and lists (\ref{subsec:resol}) can confuse common NLP tools.  The overall effect is that many subjects, verbs, and objects are misclassified and unresolved.

\begin{figure} [!tp]
	
	\centering
	\vspace{-10mm}
	\hspace*{-3mm}
	\includegraphics[width=0.50\textwidth]{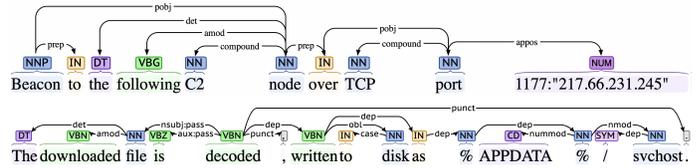}
	\vspace{-40mm}
	\caption{ POS tags and DP tree obtained from \cite{spacy} (on the top) and \cite{stanford} (on the bottom). Tags inside the boxes represented POS and tags on the arches signify the dependencies. The figure shows examples of imprecise tagging of the technical cybersecurity sentences by statistical models. }
	 		\vspace{-4mm}
	\label{fig:pos-dp}
\end{figure}

\noindent
\textbf{Relationships Extraction.} Overcoming the previous challenges can help to pinpoint the correct entities that are important in an attack description. The next step is to determine {\em “who did what to whom”}, {\em “when”} and {\em “where”} or in other words, we need to discover the relationships between process and system objects and their mapping to audit events. Current approaches related to this task, such as statistical dependency parsers are known for performance degradation on sentences drawn from domains different from that of natural English text \cite{mcclosky2010automatic, mcclosky2006reranking}. 
To resolve this issue, we need a more comprehensive approach that takes the sentence's semantics into account rather than only relying on the sentence's syntactic structure (i.e., pure use of dependency parsing, as used by \cite{liao2016acing,husari2017ttpdrill,zhu2018chainsmith}).  As we will show in detail in the next section, to solve this challenge, we use {\em Semantic Role Labeling (SRL)}, a processing model that can detect semantic relationships among entities in a sentence. 

Before continuing with the description of our approach, we provide a brief background on the NLP techniques that are used throughout the paper. 

\begin{figure*} [!t]
	\vspace{-34mm}
	\centering
	 	\hspace*{-1mm}
	\includegraphics[width=1\textwidth]{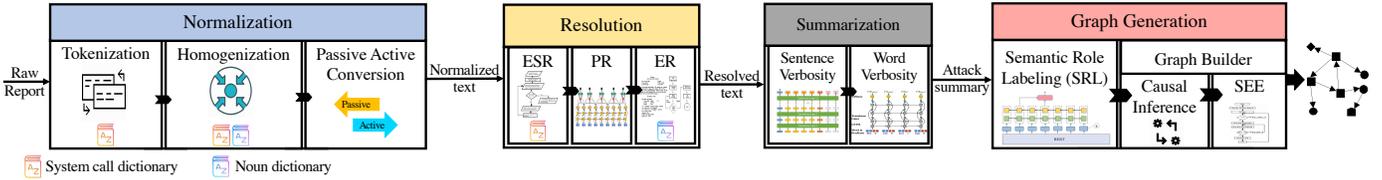}
	\vspace{-90mm}
	\caption{ An overview of EXTRACTOR architecture  }
	\label{fig:architecture}
\end{figure*}

 \subsection{NLP Background}\label{subsec:info_extraction}

\noindent\textbf{Part of Speech (POS).} POS tagging assigns a syntactic role to each word in a sentence (e.g., noun, verb, etc). In some cases, however,  POS model may fail to correctly tag words. In Figure \ref{fig:pos-dp}, adjectives \texttt{following} and \texttt{downloaded} are incorrectly tagged as {\em verb (VBG/VBN)}.

\noindent\textbf{Dependency Parsing (DP).} DP assigns grammatical connections and dependencies between words in a sentence. Example of DP tags include {\em nsubj} for sentence subjects, {\em obj} for sentence objects, etc. However, if the sentence complexity increases, DP may not be able to detect tags and the relations between words accurately. In addition, DP taggers may not be able to assign the correct tags, especially if they are not trained on context containing technical language. Common errors include tagging past participle forms as adjectives, verbs as nouns, etc. Another drawback of utilizing DP in our current problem is that the tags they produce only refer to grammatical relations, such as {\em subject, object} etc. Therefore, they cannot help in tasks that require an understanding of the semantics between different sentence components. These relationships may include temporality (when something happens), modality (how something happens), etc. In fact, a much deeper understanding is needed to accomplish our goals.

Figure \ref{fig:pos-dp} demonstrates examples of POS tags and DP trees driven by spaCy \cite{spacy} (top) and Stanford \cite{stanford} (bottom), where incorrect tags such as  {\em following} and {\em downloaded} (as discussed earlier) caused the incorrect generation of DP relations as subject and objects. Another example is the verb {\em Beacon} at the top sentence which incorrectly tagged as a proper noun (i.e., NNP). More about POS and DP  can be found at \cite{de2008stanford} and \cite{spacytags}.

\begin{table}[!t]
	\caption{ List of general arguments used in SRL (based on PropBank\cite{palmer2005proposition})	}
	\small
	\centering
	\resizebox{9cm}{!}{
	\begin{tabular}{|c|c?c|c|}
		\hline
		\textbf{Label} & \textbf{Role (argument)}           & \textbf{Label} & \textbf{Role (argument)}               \\ \hline
		ARG0           & Agent                              & ARG3           & Starting point, Benefactive, Attribute \\ \hline
		ARG1           & Patient                            & ARG4           & Ending point                           \\ \hline
		ARG2           & Instrument, Benefactive, Attribute & ARGM           & Modifier                               \\ \hline
	\end{tabular}
 }
	\label{table:srlarguments}
\end{table}

\noindent
\textbf{Semantic Role Labeling (SRL). } SRL essentially determines {\em “who did what to whom”}, {\em “when”} and {\em “where”} \cite{palmer2005proposition}. SRL is a more recent NLP technique,  which can assign semantic labels to phrases and words in a sentence, where each label specifies the {\em semantic role} that each phrase or word plays in the sentence in association with the predicate or verb of the sentence. In SRL, the tags assigned to sentence components are called {\em arguments} (denoted by ARG). Some argument examples and the corresponding semantic roles are shown in Table \ref{table:srlarguments}.

\section{Approach}\label{sec:methodology}

In a nutshell, \projname operates by performing different rounds of transformations on the text to bring it from a highly complex and potentially ambiguous form to a simpler form. This simplified text is further processed to obtain a provenance graph that can be successfully used for threat detection. An overview of \projname is shown in Figure \ref{fig:architecture}. \projname has four major components: 1) Normalization, 2) Resolution, 3) Summarization, and 4) Graph Generation. {\em Normalization} is responsible for an initial round of sentence simplification and transformation to a canonical form. {\em Resolution} resolves ambiguities in those sentences (these two components help to address CTI text complexity challenge). {\em Summarization} removes the portion of text that is not strictly related to the attack behavior, and that cannot be observed in the logs. Finally, {\em Graph Generation} is responsible for resolving the temporal and causal order among the events in the text and for building the final provenance graph (this component addresses the Relationships Extraction challenge).
Some of these components may be assisted by a set of dictionaries that contain terms related to CTI language (relying on domain-specific dictionaries of concepts is a common approach in many knowledge-based NLP systems \cite{riloff1993automatically,pfaff2015natural,DBLP:journals/corr/abs-1809-03599}). In particular, \projname uses two dictionaries. First, our system call synonym dictionary, which contains verbs representing system calls (e.g., write, fork) and their corresponding synonyms. These synonyms represent the possible verbs that can be used in CTI reports and very likely refer to a  system call.  Second, our CTI nouns dictionary contains noun phrases commonly used in CTI reports, as well as different textual representations of the same concept. The former contains 87 verbs representing system calls, while the latter holds over 1112 common noun phrases in the CTI report. Both dictionaries are depicted in Figure \ref{fig:architecture}, and will be further discussed in Section \ref{sec:implementation}.

\subsection{Normalization}\label{subsec:preprocessor}

To address the CTI text complexity challenge and maximize the accuracy of the techniques used by \projname, we must first have some canonical sentence form. We achieve this through Normalization, which is responsible for breaking long and complex sentences into shorter sentences appearing in a canonical form, which is easier to process. Intuitively, we would like each sentence to express a single action so that the subject and object of the action and the action itself be easier to identify. Normalization is comprised of Tokenization, Homogenization, and Conversion. These steps perform the detection of sentence boundaries, word homogenization, and passive-to-active verb conversion, respectively. We describe each of these steps next.

\begin{figure*} [!t]
	\centering
	\vspace{-34mm}
	\includegraphics[width=0.9\textwidth]{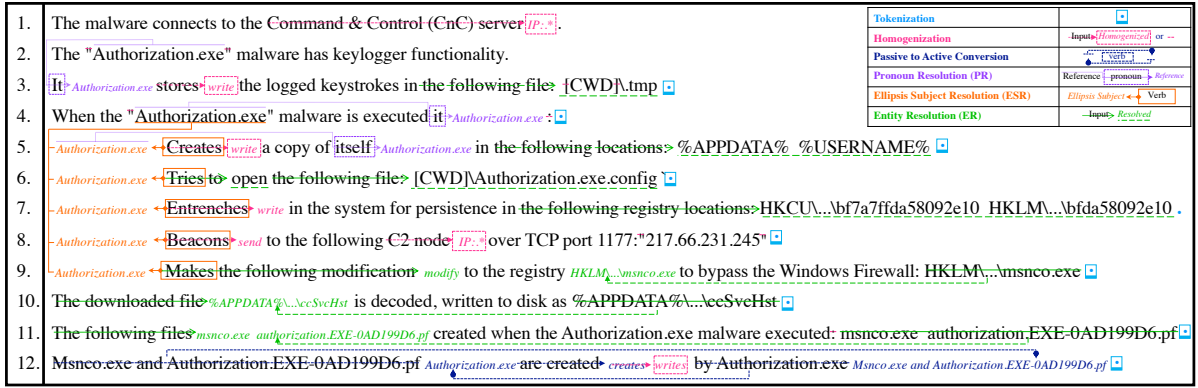}
	\vspace{-44mm}
	\caption{Transformation steps to turn a CTI report into a digestible form. A free adaptation from the njRAT \cite{solutions2013njrat}  where irrelevant sentences are removed. Lines with pointer signifies the reference and the strike-lined pointer shows the original phrase and its substituted output.  The figure best appears in color.  }
	\label{fig:running}
\end{figure*}

\noindent\textbf{Tokenization.}  
Correctly defining sentence boundaries is very important as several techniques used by \projname operate at the sentence level. 
However, existing sentence tokenizers (e.g., NLTK~\cite{loper2002nltk}) usually take only classic punctuation (\texttt{`.,!,?'}) into account when discovering sentence boundaries and perform poorly on CTI reports. In fact, in this domain, there is a high prevalence of long sentences containing multiple actions and non-standard sentence delimiters. For instance, in Microsoft threat reports, the average sentence length is almost four times higher than that of common English sentences.

To solve this problem, we design an enhanced tokenizer specialized for CTI reports. In particular, in addition to the classic sentence delimiters, our tokenizer uses new lines, bullet points, enumeration numbers, and titles and headers, as {\em sentence delimiters} to partition long sequences into sets of shorter ones. After breaking long sentences into shorter sequences of words, each short sequence is `promoted' to a sentence if it satisfies one of the following cases; 1) the sequence starts with a capitalized subject, it contains all the components necessary to form a complete sentence (subject, predicate, object), and the preceding and subsequent sequences also form complete sentences; 2) the sentence starts with a verb contained in the system calls dictionary, it contains all the components necessary to form a complete sentence minus the subject, and the preceding and subsequent sequences also form complete sentences. The latter case represents the common phenomenon (in CTI reports) of Ellipsis Subject (see Section \ref{subsec:charac}). If none of the above two cases is satisfied, we consider the sentence as an unbreakable full sentence.

As an example of this procedure, consider Figure \ref{fig:running}, which illustrates several techniques described in the paper. In this figure, the long sentence spanning lines 4-9 in Figure \ref{fig:running} is first partitioned into shorter sequences (one per line in the figure). Next, each sequence is tagged by a POS tagger and DP, and checked if it satisfies one of the two conditions above. In the figure, the sequence at line 4 satisfies the conditions of the first case, while the sequences at lines 5-9 satisfy the conditions of the second case (ellipsis subject).

The result of the tokenizer is a set of shorter sentences that is more likely to describe a single action.

\noindent\textbf{Homogenization.} 
CTI reports often contain constructs and synonyms that can introduce ambiguities and impact the final results' quality.
For example, {\tt C2, C\&C}, and {\tt Command and Control} are different representations of the same entity, while verbs like {\tt stores}, {\tt saves} may represent an action that corresponds to a {\tt write} system call. {\em Homogenization} is the process by which multiple textual representations of the same concept are replaced by the same textual representation. 

We perform {\em Homogenization} for noun phrases and verbs using two specially built dictionaries, which map different jargon and synonyms of nouns and verbs present in CTI reports to entities and actions that can be observed in audit logs. For instance, each among {\tt C2, C\&C, Command and Control} is mapped to {\tt IP:.*}, which is a wildcard representing IP addresses.  
In the same manner, we translate verbs that are synonyms with a system call inside the system call dictionary with that system call verbs.

Homogenization significantly reduces text's heterogeneity and supports our goal of providing actionable intelligence. We decide that the single word that is chosen to represent all the other words of a synset is one that is highly likely to be as a system entity that is observable in the logs or a {\em system call}.

\noindent\textbf{Conversion.} 
As the last step of text normalization, \projname converts passive voices to active. This conversion helps with discovering system subjects (processes) and system objects, as well as making causality inference more accurate, as discussed in Subsection \ref{sub:extraction}. 

To perform this conversion, we first detect passive sentences using POS and DP tagging.  This kind of sentence is predominantly represented by specific and known patterns in DP trees. For instance, consider the sentence {\tt the downloaded file is deleted by the malware}. In the DP tree, {\em is} is tagged as an auxiliary (and passive) verb, {\tt deleted} as a verb and head of the DP tree, {\tt the downloaded file} is a noun phrase that is the subject of a passive voice ({\em nsubjpass}) and {\tt by malware} is the object {\em (obj)}. Note that in some cases, the agent does not appear in a passive sentence but is implied. For instance, in line 10 in Figure \ref{fig:running}, the agent is the malware, but no references to it appear in the sentence.  
Using these patterns, \projname can detect passive sentences and distinguish between passive sentences with explicit agents and those with implicit agents. In the former case, it switches the agent and the subject, and it conjugates the passive verb to an active verb.

The final result of this step is that long sentences are transformed
into short ones in an active form, likely to express one action per sentence.

\subsection{Resolution}\label{subsec:resol}
After {\em Normalization}, {\em Resolution} 
reconciles implicit references that refer to the same entity into the actual referent.  These implicit references must be made explicit for two reasons. First, implicit references reduce the accuracy of the subsequent steps and make the final provenance graph ambiguous and imprecise. Second, audit logs contain only explicitly named entities, and every threat hunting approach cannot match system processes to pronouns and other implicit references. More thorough and fascinating discussions on such linguistic structures can be found at ~\cite{soon2001machine,ng2002improving,DBLP:journals/corr/abs-1805-11824}.

\noindent\textbf{Ellipsis Subject Resolution (ESR).} 
As discussed in Section \ref{sec:background}, ellipsis subject is a linguistic structure where a sentence's subject is not present. This kind of structure is shared in a large number of CTI reports and used for describing a sequence of actions carried out by the same actor (process or attacker)- Section \ref{sec:evaluation} presents the popularity of this phenomenon in various sources. The omitted subjects confuse the state-of-the-art NLP toolkits, thus resulting in the loss of the narrative sequence and the story relationships (subject and object of an action). All the actions described in lines 5-9 in Figure \ref{fig:running} are examples of ellipsis subjects.

To address this problem, we developed an Ellipsis Subject Resolver (ESR) module. This module utilizes POS and DP parsing along with the system calls dictionary. The first step in resolving this problem is the detection of sentences with missing subjects. This step uses POS and DP together with the system calls dictionary, as was described in the discussion about the Tokenizer (Subsection \ref{subsec:preprocessor}). Once this kind of sentence is detected, ESR builds a list of candidate subjects among the entities appearing in the sentences preceding the current sentence. Next, the module picks the most probable candidates from the list based on the distance (computed as the number of sentences) of that candidate from the sentence with the missing subject. In particular, the closer candidate has a higher probability of being picked. For instance, in Figure \ref{fig:running}, the subject is missing in the sentences in lines 5-9. The ESR module detects the subjects and other objects in the previous sentences, and it chooses the pronoun {\em it}
occurring right before the colon as the subject.

\noindent\textbf{Pronoun Resolution (PR).} Pronoun resolution is the process by which pronouns are mapped and substituted to the antecedent entities that they refer to. Processing documents (building a provenance graph) without PR can result in the appearance of several nodes (i.e., pronouns) for a single entity. For instance, in Figure \ref{fig:running}, the pronoun \texttt{it} and \texttt{itself} in lines 3 and 5 should be replaced with the actual subject \texttt{Authorization.exe}.

To resolve pronouns, we adapt a popular coreference resolution model, NeuralCoref \cite{NeuralCoref}. We noticed that this model works best in resolving pronouns in the CTI reports domain, especially after the previous steps of ESR, and Tokenization. 
Figure \ref{fig:running}, lines 4, 5, and 6 demonstrate the resolved pronouns (i.e., it and itself) and their corresponding reference (\texttt{Autorization.exe}).

\noindent\textbf{Entity Resolution (ER)}. Entity resolution is the process by which noun or verb phrases that refer to another entity inside the same sentence are substituted by that entity or are eliminated as redundant. This is a vast task to perform in general, however, we point out that we are interested only in extracting {\em actionable} information and, therefore, can focus on performing ER only on entities and actions that are likely to appear in audit logs. In fact, from a preliminary observation of a large number of CTI reports, we noticed that redundancies among those entities and actions appear under mainly three distinct linguistic forms:

\noindent{\em Anaphora}. An anaphora is the use of a word or pronoun to refer back to another word or phrase that was previously used in the sentence to avoid repetition. For instance, in line 11 of Figure \ref{fig:running}, {\tt The following files} refers to {\tt mscno.exe authorization.EXE-0AD199D6.pf}. This form is prevalent in CTI reports, where it is used to describe lists of entities participating in some common action. 

\noindent {\em Nominalization}. This is a form where an auxiliary verb is used together with a noun in place of a verb. For instance, {\em makes a modification} in place of {\em modifies}. This form is often used with actions that represent system calls. In particular, it appears approximately 3524 times in the TrendMicro and 1261 times in Microsoft blogs. Another similar form related to system calls appears as an auxiliary verb followed by an actual verb related to a system call, e.g., {\em tries to open} instead of {\em opens} in Figure \ref{fig:running}.

To resolve these cases, we use a combination of POS tagging and DP with domain knowledge contained in  {\em CTI nouns} dictionary or in a corpus of common phrases appearing in each case (e.g., {\em the following files} is a common anaphora). In particular, if one of the above three forms is detected in the text, we retrieve the DP and POS tags of the other words in the vicinity of that form and check that they follow specific patterns. In particular, for anaphoras, we check that a list of noun phrases follows the main sentence where the anaphora appears and replace the anaphora with the noun phrases. For nominalizations, we check that the noun present in the corpus is the object of a preceding auxiliary verb and replace that noun with its verb form (e.g., {\tt makes the modification} $\rightarrow$ {\tt modifies}). For auxiliary verbs, we detect if an infinitive form precedes a verb that may represent a system call and replace the whole phrase with the actual verb ({\tt tries to open} $\rightarrow$ {\tt open}).

\begin{figure*} [ht]
	\vspace{-18mm}
	\centering
	\includegraphics[width=0.9\textwidth]{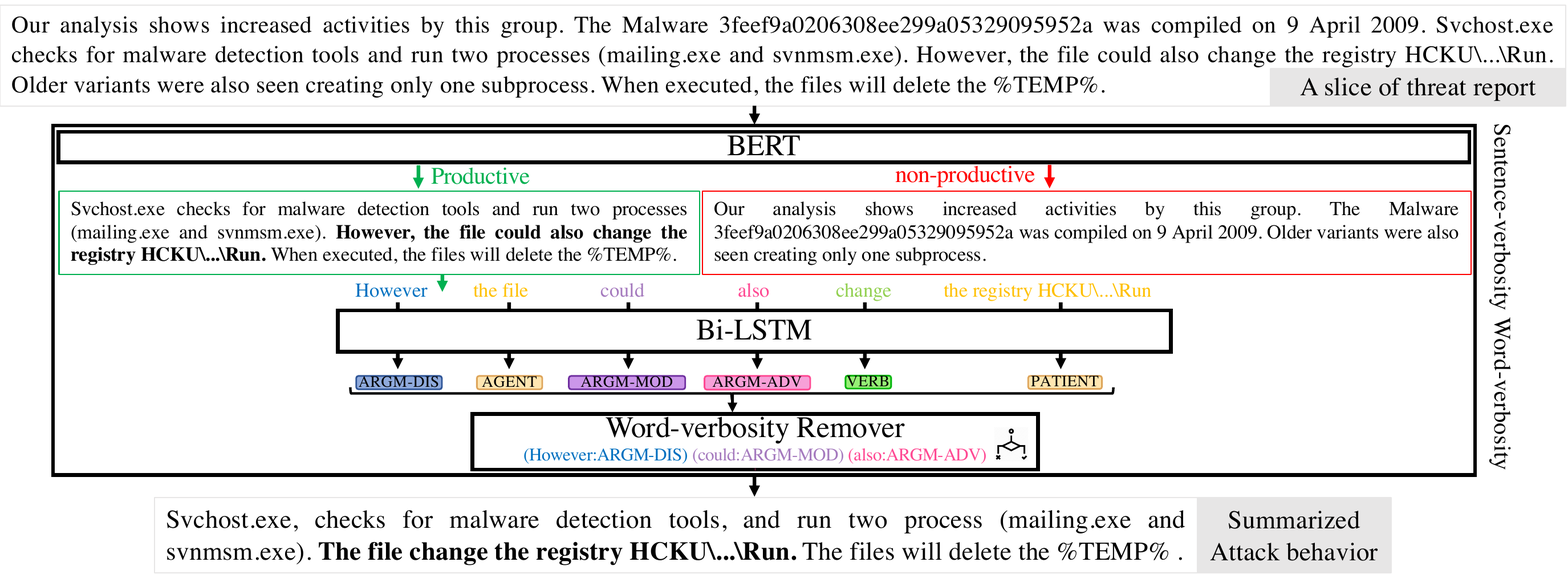}
	\vspace{-50mm}
	\caption{ 
		The architecture of the Text summarizer. The module reduces the verbosity using BERT, BiLSTM, word-verbosity remover. }
	\label{fig:summarizer}
	
\end{figure*}

After the {\em Resolution} step is completed, the text consists of sentences having explicit subjects, objects, and verbs. The amount of text is also reduced somewhat by the ER module. However, the major text reduction step is executed after {\em Resolution} and is described next.

\subsection{Text Summarization} \label{subsec:ats}

To reduce verbosity and obtain a concise description of the attack behavior that can be directly used to detect the attack, a significant amount of superfluous text must be removed.
Ideally, only the sentences that describe actions that may be observed in the audit logs should be preserved. To do this, \projname must understand which sentences strictly describe attack behavior and which sentences do not. Previous related work uses {\em topic classification}  \cite{liao2016acing,husari2017ttpdrill} to identify {\em topic}-related context among out-of-domain contexts (e.g., advertisement text versus technical text).  While these approaches are successful in separating irrelevant content (such as ads) from technical content, they are not powerful enough to separate the latter into {\em behavioral} content that describes observable attack actions from other ``technical'' content, which serves as an introduction or context description. We refer to this problem as {\em sentence verbosity}. An example of {\em sentence verbosity} is shown in Figure \ref{fig:summarizer}. In the figure, the text of the report is shown at the top. The sentences in the box on the top left corner, labeled by {\em Productive}, contain a description of the malware's actual behavior, which can be observed in audit logs, and which can, therefore, be useful for detection. The sentences on the top right corner, labeled by {\em non-productive} contain the complementary description of the malware but no actions that can be observed in audit logs. Even though the two text portions are technical in nature and about the same {\em topic}, we are only interested in the {\em productive} text and want to remove the {\em non-productive} one.

Another problem that needs to be solved is what we refer to as {\em word verbosity}. In particular, inside each sentence, there usually appear word constructs, such as adverbial and adjectival phrases, which do not contribute to the behavior description and can be safely removed (e.g., {\tt However}, {\tt could}, and {\tt also} in the figure).

To deal with these problems, we design a two-step approach. This approach is shown in Figure \ref{fig:summarizer} and is composed of a BERT classifier, which deals with {\em sentence verbosity}, and a BiLSTM network, which deals with {\em word verbosity}.

\noindent
\textbf{Sentence Verbosity}. To distinguish sentences that describe actual threat behavior from the ones that do not represent threat behavior, we need to go beyond {\em topic classification} and have a deeper understanding of the text. Intuitively, {\em productive} sentences express more ``direct'' connections between the subject and the object than the other sentences. Thus, to classify these connections, a linguistic model of the text must build a finer-grained representation of the words' context. 

Currently, one of the best models to build such fine-grained representation is BERT (Bidirectional Encoder Representations from Transformer) \cite{bert}. Unlike Word2Vec \cite{w2v} and GloVe \cite{glove} word representations, BERT builds contextual representations of the words that take into account both the text before a word and the text after a word. In other words, BERT considers the context surrounding each word. In addition, BERT learns embeddings for subwords, that is sub-components of a word derived from stemming. This allows the model to more effectively deal with out-of-vocabulary words.
In general, this capability is beneficial for the technical cybersecurity documents, which may contain lexically complex phrases that do not appear in BERT's training set. As a result, BERT can classify sentences into productive and non-productive much better than other approaches. In particular,  we labeled 8,000 threat sentences under two classes of productive and non-productive, and trained BERT on this set. The results are promising and are shown in detail in Section \ref{sec:evaluation}.

\noindent
\textbf{Word Verbosity}. The second step of the Text Summarizer removes unnecessary words from the {\em productive} sentences that it receives as input from BERT. It is composed of two phases, a BiLSTM network that derives the semantic roles of the sentence components and a word remover phase. We found that BiLSTM works best for this purpose since it can handle long-distance dependencies that appear in technical documents. 

After a sentence is processed by a BiLSTM network, its components are tagged as {\em Agent}, {\em Patient}, and {\em Action}, and other types of arguments (e.g., in Figure \ref{fig:summarizer} the word {\em However} is labeled as {ARGM-DIS}, a discourse marker that connects a sentence to a preceding sentence). In the next phase, the unnecessary sentence components are removed. In theory, this can be done only by keeping the {\em Agent}, {\em Action}, and {\em Patient} components of the sentence. However, in certain cases, this approach would remove important information. For instance, a sentence such as {\em when malware.exe is executed} may be labeled as ARGM-TMP (a temporal marker), and removed by a naive approach, leading to the removal of an important part of the attack. To make this second phase more precise and not remove sentence components that may contain important objects like {\em malware.exe}, we use the {\em System Entity Extractor (SEE)} which will be introduced in Section \ref{subsub:gb}. In particular, a sentence component that is tagged for removal will be removed if it does not contain any entities that can be generated by the rules of the {\em SEE} component. 

Text Summarization is one of the central components of \projname. It is responsible for greatly reducing the text's complexity and quantity while keeping the most important sentences that describe observable behavior.

\subsection{Graph Extraction}\label{sub:extraction}
After the previous steps, the resulting text is in a form where the system subjects (e.g., process), objects (e.g., file, socket), and actions (e.g., exec) are explicit, well ordered, and a large part of the superfluous text is eliminated. In this last step, \projname addresses the challenge of relationship extraction to extract a provenance graph from the simplified text. 

Even though the text at this step is very simple, a naive graph extraction that assigns nodes to subjects and objects, and edges to the verbs would create ambiguous and large graphs. This is because several roles and relationships between subjects and objects may be expressed in the same sentence. To deal with this problem, we use Semantic Role Labeling (SRL) and a set of rules to extract the causality relations and directions of information flow. These are described next. 

\begin{figure} [!t]
	\vspace{-11mm}
	\centering
	\hspace*{-5mm}
	\includegraphics[width=0.53\textwidth]{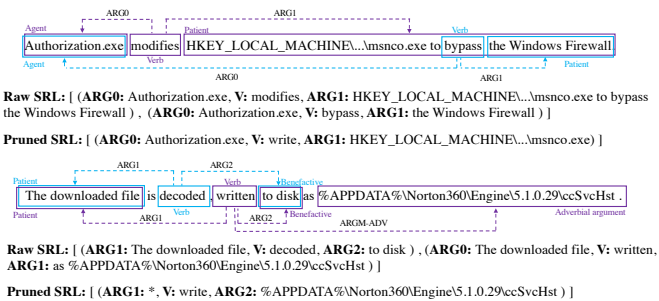}
	\vspace{-28mm}
	\caption{Examples of semantic roles and relations. Roles are generated according to the PropBank annotations. Words on the arch present the labels and roles are signified by words on the top/bottom of the rectangles. }
	\label{fig:srl}
\end{figure}

\subsubsection{Semantic Role Labeling (SRL)}
As mentioned in Section \ref{sec:background}, SRL is a technique that discovers the semantic roles in the sentence. 
To give an intuitive overview of the power of SRL, consider the two examples in Figure \ref{fig:srl}, one in the active and one in the passive form. SRL is able to extract two roles from each sentence (denoted by {\em Raw SRL}) and understand which noun is the patient (that is the one the action falls on, denoted by ARG1) and which is the agent (the noun carrying the action, denoted by ARG0). For the purpose of our discussions, a {\em SRL role} can be thought of as an action. SRL is, therefore, able to correctly associate each component in a sentence with a semantic tag.

\projname considers all the possible arguments related to a verb detected by SRL as potential subjects or objects of the attack and then prunes out those which are not system entities. For the pruning process, we use the {\em SEE} module (see Section \ref{subsub:gb}), which detects possible system entity names (e.g., file or process names, IP, and registry keys). In particular, the {\em SEE} module analyzes each node and prunes out the whole node or part of the node name that does not match one of the regular expressions or application names. The result of the pruning in Figure \ref{fig:srl} is denoted by {\em Pruned SRL}.

\noindent
\textbf{Actions to System Calls Transformation.}
After SRL, \projname performs a second {\em homogenization} step over the verb roles detected by the SRL module. This step is necessary to correct eventual errors due to POS and DP taggers' inefficiency, which might allow verbs to slip through and remain in their original form (untranslated to system calls). 
After this step, we prune away those roles created by SRL that do not represent a system call action. 
For instance, after this second pruning step, the second role related to the verb {\tt bypass} in the top half of Figure \ref{fig:srl} is pruned out.

\subsubsection{Graph Builder (GB)} \label{subsub:gb}
The final step of our approach is to construct the graph from the output of SRL.  The {\em GB} operates in two steps. First, it {\em merges} the {\em SRL arguments} that have the same text into the same node, and using {\em SEE} prunes out words that are not system entities. Next, {\em GB} builds the graph using: 

\begin{enumerate}[noitemsep,topsep=0pt]
	\item {\em Node-edge-node triples. } For every sentence, if it has at least three roles including a verb role (a system call representation as a connector) and two nodes, GB generates the edge and node pairs. 
	\item {\em Edge direction.} \projname determines the direction of the edges by using a small map of edge directions associated with the system calls dictionary. This is discussed in more detail the {\em Causal Inference} paragraph later in this section. 
	
\end{enumerate}

\noindent\textbf{System Entity Extractor (SEE).} We developed the {\em SEE} module to extract concise nodes that represent system entities from the roles generated by SRL, and to prune out the futile part of speeches that cannot constitute possible system entities. SEE detects possible system entity names (e.g., file or process names, IP, registry keys) using over 32 different regular expressions and a database of application names or well-known processes. In particular, the {\em SEE} module analyzes each noun phrase and prunes out the whole phrase or the part of it that does not match one of the regular expressions or application names. For example, in the sentence, \texttt{The malware deleted the regex.exe.}, {\em SEE} prunes out the (ARG0) into asterisk (* - which in query processing systems will be inferred as any) and turns (ARG1) into regex.exe.
This step is essential to have concise and accurate names for the system entities that can be used to search the audit logs for threat detection. This module also defines the shape of the nodes in the final graph, where the rectangle, oval, pentagon, and diamond represent the file, process, registry, and socket, respectively. 

\noindent\textbf{Causal Inference.} 
This step determines the correct direction of the edges in the graph to represent causality and information flow among nodes. To infer this direction, it uses a mapping of system calls to the direction of system flows. 
The mapping contains entries that associate with each system call the direction of the edge between the subject and the object (e.g., for the {\em send}  system call the flow goes from subject to object, while for the {\em recv} it goes from object to subject).
Besides, this step addresses negated verbs, which may appear in CTI reports. In fact, in the casual inference, we detect the negation using SRL tags and purge the negated roles if there are no conditional clauses that influence the role. For instance, \texttt{svchost.exe does not create explorer.exe} will be purged as no conditional clauses influences this sentence.

The output of the last step is a provenance graph that clearly shows the entities that participate in and are affected by the attack as nodes, as well as the system calls connecting them as edges. An example of such a graph related to the running example is shown in Figure \ref{fig:background}.

\section{Implementation}\label{sec:implementation}

In this section, we briefly describe some additional implementation details, tools, and techniques used by \projname.

\noindent
\textbf{NLP toolkits.} We used a combination of various state of the art NLP toolkits to implement our approach. These include the {\em spaCy} POS and DP tagger, {\em NLTK}, and {\em Stanford}~\cite{spacy,stanford,loper2002nltk}. 
We use {\em SpaCy} in {\em Tokenization, Homogenization, Resolution}, and  {\em Passive to Active Conversion} steps to determine the POS and DP tags of the different sentence components. In particular, we used the large pre-trained statistical model version 2 \cite{spacylg} of spaCy as the model outperforms the other statistical models in dealing with CTI reports.

\noindent\textbf{Tokenization}. Our sentence tokenizer is built on top of the NLTK sentence tokenizer. NLTK (Natural Language ToolKit) is a common NLP toolkit, containing several libraries and statistical natural language processing developed for the English language. We chose NLTK because we found that it  works better than others (spaCy, Stanford) and it is more consistent in dealing with text in the CTI domain.

\noindent\textbf{Text Summarization}.  We used a 12 hidden layer BERT \cite{bert} to discern the {\em productive} sentences from the {\em non-productive} ones.  To train our model, we used 8,000 labeled sentences. To understand the words' roles in the text summarizer, we used a re-implementation of a deep BiLSTM model \cite{he2017deep}. Since the model was not fine-tuned to handle cybersecurity sentences, we trained the model using 3,000 manually labeled sentences. 

\noindent\textbf{SRL}.
To implement SRL, we use the method described on \cite{shi2019simple}, deployed by \cite{gardner2018allennlp}, which is becoming increasingly popular in the NLP area.  To adopt the system and receive more precise output in the cybersecurity domain, we further retrain the model with 2,000 cybersecurity sentences related to the areas in which we notice that the system fails to predict the roles properly. 
For further completeness, we evaluated and presented the result of our retraining ( see Section \ref{subsec:ETS}).

\begin{table}
	\caption{System call verbs and their corresponding  synonyms in system call dictionary.}
	\begin{adjustbox}{width=1\linewidth}
		\small
		\begin{tabular}{|c|c|} 
			\hline
			\textbf{System call}  & \textbf{Synonyms}                                                                                                                                                                                                            \\ 
			\hline
			\textbf{Write}                 & \begin{tabular}[c]{@{}c@{}}write, form , entrench, place, exfiltrate, deploy, implant, drop, install, \\ putfile, compose, create, copy, save, add, modify, append, create\end{tabular}  \\ 
			\hline
			\textbf{Read}                  & \begin{tabular}[c]{@{}c@{}}survey, read, gather, download, navigate, locate, get, \\ acquire, check, detect, record, exfiltrate , extract, obtain \end{tabular}                                  \\ 
			\hline
			\textbf{Unlink}                & unlink, delete, clear, remove, erase, wipe, purge, expunge                                                                                                                                                   \\ 
			\hline
			\textbf{Send}                  & send, transfer, post, postsinformation, move, transmit, deliver, push, redirect                                                                                                                            \\ 
			\hline
			\textbf{Receive}               & receive, accept, take, get, collect                                                                                                                                                                                \\ 
			\hline
			\textbf{Connect}               & connect, click, browse, portscan, communicate                                                                                                                                                                      \\ 
			\hline
			\textbf{Fork}                  & fork , clone, spawn, issue, set                                                                                                                                                                                    \\ 
			\hline
			\textbf{EXEC}                  & \begin{tabular}[c]{@{}c@{}}use, execute, executed, run, launch, call, \\ perform, list, invoke, inject, open, target, resume \end{tabular}                                                         \\ 
			\hline
			\textbf{Exit}                  &  exit, terminate, stop, end, finish, break off, abort, conclude                                                                                                                                              \\ 
			\hline
			\textbf{MMAP}                  & allocate, assign                                                                                                                                                                                                         \\
			\hline
		\end{tabular}
	\end{adjustbox}
	\label{synnet}
\end{table}

\noindent\textbf{Dataset and Dictionary Construction.}
To build dictionaries and our datasets, we used our pool of CTI reports scraped from various sources. We used different sources of namely APT report repository \cite{aptreportsgithub}, Microsoft Threat Center \cite{msthreat}, Symantec Security Center\cite{symanteccent}, Threat Encyclopedia \cite{trendmicroenc}, and Virus Radar \cite{virusradar} to ensure the diversity and coverage.

For our text summarizer, we annotate a balanced dataset of 8,000 sentences sampled from various sources and annotated with two categories: productive and non-productive sentences. In total, 3,800 sentences are annotated as productive, 4,200 messages are annotated as non-productive sentences. We split our dataset into 4,800 sentences for training, 1,600 sentences for validation, and 1,600 sentences for the test. We used distinct sets for test and evaluation.

We perform annotation in an iterative fashion, and three subject matter experts were involved in the annotation of our datasets. We request each annotator to annotate the collected data into one of the two categories of productive and non-productive. Then, on several discussions with the annotators, we discuss and clarify the notion of the attack behavior (i.e., productive) versus the rest (i.e., non-productive) to ensure the understanding of attack behavior is accurate.
Following prior guidelines and studies (i.e., \cite{fort2016collaborative} and \cite{d2015influence}), the annotation task begins in an iterative fashion. In each round, 200 messages are assigned, and disagreements are discussed with each annotator. After each round of discussions, 100\% inter-annotator agreement (IAA) is achieved as measured by Cohens kappa coefficient. After three initial rounds of annotations, the annotators are assigned the remaining 7,400 sentences, where an IAA of 91\% is obtained. The final round of disagreements are discussed, and labels are finalized by one of the authors of this paper. 

An alternative solution to translate the verb phrases into the corresponding system call is to use tools like WordNet \cite{miller1995wordnet} and Thesaurus \cite{thesaurus} (researchers like \cite{husari2017ttpdrill} have previously used this kind of approach). However, we noticed that we could achieve better results by creating a simple though effective dictionary.  To build our dictionaries, similar to the process of annotating our dataset, we worked with a team of three security experts in an iterative fashion. The members were involved in reviewing and annotating 3000 randomly selected technical threat reports from various sources over a period of one year. Then, in an iterative fashion, the extracted phrases and their corresponding synonyms have been discussed and agreed.  Similarly, the system calls dictionary is derived from WordNet \cite{miller1995wordnet} and Thesaurus \cite{thesaurus}. These synonyms have been extracted and discussed in several discussions to assure the quality. Tables \ref{synnet} presents this dictionary. Also, Table \ref{noun}, in Appendix, represents examples of the noun dictionary.

\section{Evaluation}\label{sec:evaluation}

To evaluate \projname, we designed three experiments, each performed on CTI reports with distinct writing styles. In the first experiment (\textsection\ref{subsec:publicctireport}), \projname generates graphs from a set of public CTI reports describing real-world incidents. In the second experiment (\textsection\ref{subsec:darpa}), \projname builds graphs from the descriptions of attacks in the DARPA Transparent Computing program \cite{darpa} evaluations. Finally, in a large scale experiment (\textsection\ref{subsec:largscale}), \projname processed 4,100 unstructured CTI reports from Microsoft Security Intelligence \cite{msthreat} and 11,600 reports from TrendMicro  \cite{trendmicroenc} to extract  provenance graphs as further discussed in Section \ref{subsec:largscale}.

In the first two experiments, we evaluate \projname in two distinct ways: (1) We measure \projname's capability in capturing all relevant attack behaviors using the ground truth present in the reports. To this end, we report precision, recall, and F1-score. (2) To demonstrate the usefulness of \projname in supporting threat hunting, we use a threat hunting system, POIROT  \cite{milajerdi2019poirot}, with the graphs generated by \projname. Finally, to evaluate the scalability of our approach, we perform a large scale experiment, which is discussed in Section \ref{subsec:largscale}.

\textit{Threat Hunting.} To evaluate the usefulness of the graphs generated by \projname for threat detection, we used POIROT system \cite{milajerdi2019poirot}. This system takes as input a small provenance graph, called {\em query graph}, representing attack events, and searches for embeddings of that graph in a much larger provenance graph built from the audit logs of the systems under attack. The query graphs in POIROT are manually  built by experts after reading CTI reports and represent the attack activities described in those reports. In our evaluation, we use the same CTI reports to automatically build graphs  with EXTRACTOR and use those graphs as query graphs for POIROT. In this way, we compare graphs built by human experts and graphs built by EXTRACTOR and the usefulness of both kinds of graphs to detect threats. 
We define an operation of $P(G_1 , G_2) = S $, where $G_1$ represents the graph built by EXTRACTOR, and $G_2$ represents the larger provenance graph representing the audit logs of the systems under attack. Next, we use POIROT to search for $G_1$ within $G_2$  and retrieve the similarity score $S$. If $S$ is bigger than the POIROT threshold ($t\approx0.3$), then $G_1$ is successfully located in $G_2$,  indicating a successful detection of a threat. Otherwise, no attack has been detected. For more details on POIROT, refer to \cite{milajerdi2019poirot}.

In all the experiments, we measure \projname's false positive and false negative edges. By false positive edges, we refer to the edges included in the extracted graph, which do not represent attack activities. By false negative edges, we refer to edges that should have been included in the extracted graph. We point out that these notions of false positive and false negative edges refer only to the presence (or lack thereof) of nodes and edges in the final graph and not to the actual detection of the threat using that final graph. In fact, many detection tools might be able to use a small set of nodes and edges as IOCs. As a specific example, the tool we used in this paper, POIROT, employs approximate graph matching using graphs with extraneous or missing edges~\cite{milajerdi2019poirot}, and is robust to a certain degree of false-positive and false-negative edges.  

\begin{table}
	\caption{Characteristics (nodes |V| and edges |E|) of the \projname vs. manual graph and results of threat detection in  CTI reports, score $P(G_1 , G_2)$ and detection outcome DO.}
\centering
	\begin{adjustbox}{width=1\linewidth}
\begin{tabular}{|c|c|c|c|c|c|c|} 
\hline
\multirow{2}{*}{ \textbf{Scenario} } & \multicolumn{2}{c|}{\textbf{Manual} } & \multicolumn{2}{c|}{\textbf{EXTRACTOR} } &
\textbf{\multirow{2}{*}{ \textbf{Score}}}
& \multirow{2}{*}{\textbf{Do}}  \\ 
\cline{2-5}
                                     & \textbf{|V|}  & \textbf{|E|}              & \textbf{|V|}  & \textbf{|E|}                 &                                                                                              &                                                                                                \\ 
\hline
njRAT \cite{solutions2013njrat}  - fig. \ref{fig:njrat}                                & 14             & 14                         & 32             & 32                            & 0.4                                                                                          &    \checkmark                                                                                             \\ 
\hline
Carbanak  \cite{carabanak}   - fig. \ref{fig:all-5}-(a)                           & 10             & 10                         & 22             & 31                            & 0.4                                                                                          &    \checkmark                                                                                             \\ 
\hline

HawkEye    \cite{fortinet}  -  fig. \ref{fig:all-5}-(b)                          & 17             & 34                         & 29             & 31                            & 0.4                                                                                          &    \checkmark                                                                                             \\ 
\hline
DeputyDog   \cite{deputydog}   - fig. \ref{fig:all-5}-(c)                          & 5              & 4                          & 11             & 12                            & 0.4                                                                                          &    \checkmark                                                                                             \\
\hline

DustySky     \cite{DustySky}  - fig. \ref{fig:all-5}-(d)                        & 9              & 10                         & 12             & 21                            & 0.6                                                                                          &    \checkmark                                                                                             \\ 
\hline
Uroburos    \cite{uroburos} -  fig. \ref{fig:all-5}-(e)                        & 12             & 15                         & 19             & 23                            & 0.5                                                                                          &   \checkmark                                                                                              \\ 
\hline

\end{tabular}
	\end{adjustbox}

\label{table:aptreports}
 \vspace{-2mm}
\end{table}

\subsection{Evaluation on Public CTI reports.}  \label{subsec:publicctireport}

In the first set of experiments, we evaluate \projname using public CTI reports. For comparison purposes, we choose the same reports chosen by the authors of POIROT  \cite{milajerdi2019poirot}. 
This experiment allows us to 1) compare the graphs generated by \projname and the graphs generated manually by the authors of POIROT, and 2) use POIROT to perform threat hunting using the graphs generated automatically by \projname and see if the attack is successfully detected. In this experiment, the audit logs contain events generated by benign activities and events generated by executing the malware instances and the same attack activities described in the CTI reports in a controlled and isolated environment, as described by the authors of POIROT~\cite{milajerdi2019poirot}.  Table \ref{table:malwares}, in the Appendix, provides additional details about each malware sample. 

Table \ref{table:aptreports} represents the characteristics of \projname graphs versus manual graphs and the result of threat detection in these public CTI reports. The first column shows the malware name, the reference to the CTI report, and the reference to the extracted provenance graph figure. The next four columns show the number of nodes ($V(G)$) and edges ($E(G)$) of the graphs manually drawn by the POIROT authors and the ones automatically generated by \projname. As can be seen, the numbers of nodes and edges extracted by \projname are comparable with the ones built manually. The main reasons for the difference in the number of nodes and edges are due to  1) the use of wildcards in the manual graphs (e.g., the use of {\em C=*.tmp} in manual instead of {\em [CWD]\textbackslash.tmp} and {\em C:\textbackslash Extracted\textbackslash.tmp}), 2) nodes and edges that are picked by \projname but are not presented in the manual graph (e.g., {\em 2: exec} - {\em 10: exec}), as the human has abstracted these details away. 
Finally, columns six and seven present the results of threat hunting, which was obtained by conducting these malware attacks in the presence of suitable benign activities, and collecting the audit records. An approximate matching algorithm \cite{milajerdi2019poirot} was used to match the \projname-generated graph inside a larger provenance graph generated from the audit logs of the systems under attack. In all scenarios, our detection score surpassed the detection threshold ($t\approx0.3$), and the attack was detected successfully. In summary, through this experiment, we can conclude that the \projname-generated graphs are as useful as human-generated graphs in threat detection. 

\begin{table}
	\caption{Precision, recall, and F1-score of graphs generated from the CTI reports, calculated by comparing edges in the automated graph against those in the CTI report ground truths.}
	\centering
	\begin{adjustbox}{width=0.7\linewidth}
	\begin{tabular}{|c|c|c|c|} 
		\hline
		\textbf{Scenario}  & \textbf{Precision} & \textbf{Recall} & \textbf{F1-Score}  \\ 
		\hline
		njRAT               & 0.90               & 1               & 0.95               \\ 
		\hline
		Carbanak           & 0.87               & 1               & 0.93               \\ 
		\hline
		Uroburos            & 0.85               & 0.96            & 0.90               \\ 
		\hline
		DustySky            & 0.85               & 0.94            & 0.90                \\ 
		\hline
		HawkEye             & 0.93               & 0.93            & 0.93               \\ 
		\hline
		DeputyDog           & 1                  & 0.92            & 0.96               \\
		\hline
	\end{tabular}
\end{adjustbox}
	\label{table:CTIFPFN}

\end{table}

\begin{figure*} [!t]
	\centering
 	\includegraphics[width=0.80\textwidth]{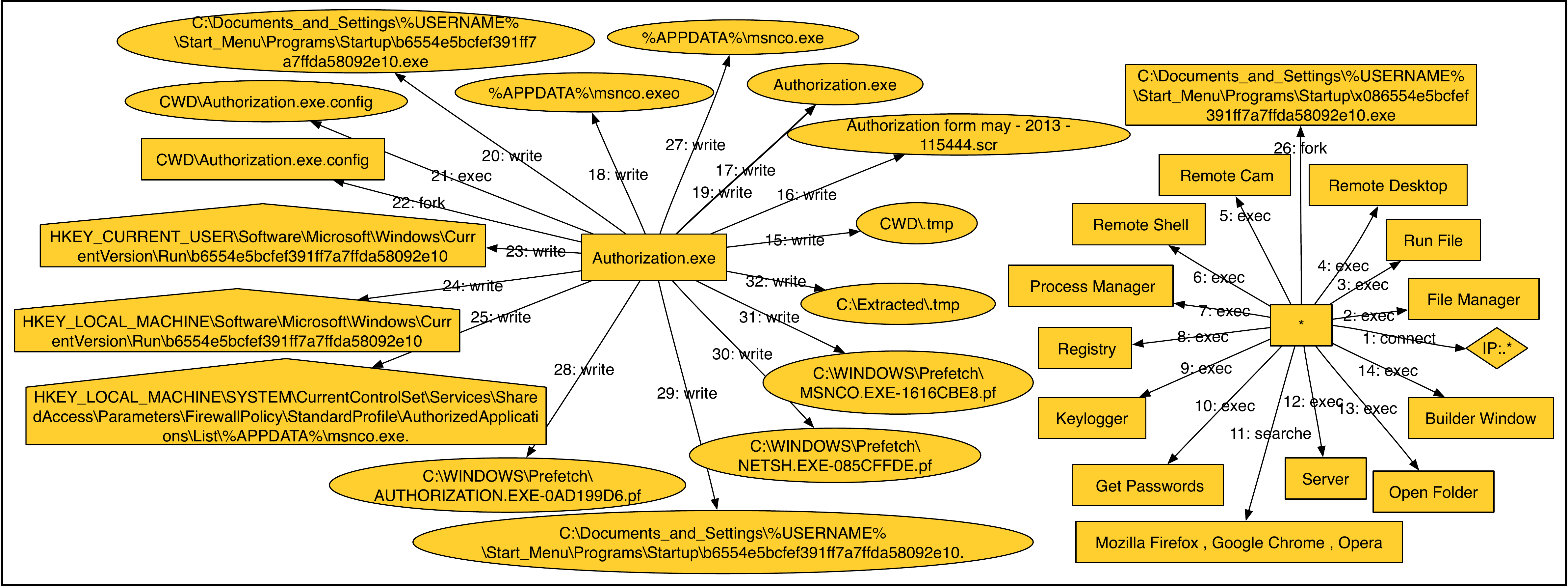}
	\caption{Graph generated by EXTRACTOR from njRAT \cite{solutions2013njrat} }
	\label{fig:njrat}
	\vspace{-1mm}
\end{figure*}

 While the results of the threat detection (Table \ref{table:aptreports}) using \projname's graphs confirm \projname's capability in capturing relevant attack behavior, to further evaluate the performance of \projname and to measure its ability in capturing all relevant attack behaviors, we report precision, recall, and F1-score (Table \ref{table:CTIFPFN}).  For this evaluation, we use the reports themselves as the ground truth and check if the activities captured in the graph are present or not in the report. We do not use the graphs generated by the experts in POIROT as ground truth, since many of those graphs contain wildcard nodes representing sets of processes. Table \ref{table:CTIFPFN} presents the performance of \projname. As shown in Table \ref{table:CTIFPFN}, \projname successfully captured attack behavior from the reports (with an average F-1 score 93\%).  However, as expected, due to language complexities, \projname yields a small number of false positives and false negatives. Sometimes this is due to inverted edges, but more often this is due to \projname not fully resolving some ambiguities or not detecting some entities in the text. For instance, in Uroburos pairs \texttt{credprov.tlb, load, explorer.exe} and \texttt{*, fork/exec, winview.ocx} are spurious nodes and edges (see  Figure \ref{fig:all-5}-(e) in the Appendix). However, these did not impact threat detection (Table \ref{table:aptreports}). 

\noindent
\textbf{Attack Descriptions.}
Figure \ref{fig:njrat} shows the graph generated by \projname from \cite{solutions2013njrat}, where the malware modifies several registry components and writes to several files. The divisions into left and right subgraphs in the figure reflects the report's structure, where it describes the actions performed by malware using various processes {\em (authorization.exe, *)}. 
Also, Figure \ref{fig:all-5}, in the Appendix, presents other graphs generated from public CTI reports, discussed at \cite{milajerdi2019poirot}, except OceanLotus \cite{OceanLotus}, in which the attack behavior is described in a figure rather than natural language description. 

\begin{table}[]
	\caption{Characteristics (nodes |V| and edges |E|) of \projname vs. manual graphs and results of threat detection in DARPA TC campaigns, score $P(G_1 , G_2)$ and detection outcome DO.}
	\small
	\centering
	\begin{adjustbox}{width=1\linewidth}
		\begin{tabular}{|c|c|c|c|c|c|c|c|} 
\hline
\multirow{2}{*}{ \textbf{Scenario} } & \multirow{2}{*}{\begin{tabular}[c]{@{}c@{}}\textbf{Number of }\\ \textbf{sentences} \end{tabular}} & \multicolumn{2}{c|}{\textbf{Manual} } & \multicolumn{2}{c|}{\textbf{ EXTRACTOR} } & 
\textbf{\multirow{2}{*}{ \textbf{score}}}
& \multirow{2}{*}{\textbf{Do}}  \\ 
\cline{3-6}
                                     &                                                                                                    & \textbf{|V| }  & \textbf{|E|}             & \textbf{|V|}  & \textbf{|E|}                  & \multicolumn{1}{c|}{}                                                                                             &                                                                                                 \\ 
\hline
Simple APT                    & 8                                                                                                  & 15              & 17                        & 13             & 13                             &    1.0                                                                                                               &  \checkmark                                                                                                \\ 
\hline
Micro APT                     & 9                                                                                                  & 13              & 15                        & 15             & 17                             &   0.9                                                                                                                &  \checkmark                                                                                                \\ 
\hline
Drakon APT                    & 10                                                                                                 & 10              & 14                        & 14             & 11                             &    0.9                                                                                                               &   \checkmark                                                                                               \\ 
\hline
GatherApp                   & 8                                                                                                  & 7               & 10                        & 8              & 8                              &    0.8                                                                                                               &   \checkmark                                                                                               \\ 
\hline
HelloWorld                  & 8                                                                                                  & 7               & 10                        & 8              & 8                              &   0.9                                                                                                                &   \checkmark                                                                                               \\ 
\hline
GatherApps                   & 8                                                                                                  & 14              & 14                        & 13             & 12                             &     0.8                                                                                                              &    \checkmark                                                                                              \\ 
\hline
Webshell                    & 9                                                                                                  & 7               & 9                         & 12             & 8                              & 0.6                                                                                                                  &      \checkmark                                                                                            \\ 
\hline
Metasploit                  & 9                                                                                                  & 21              & 22                        & 15             & 11                             &    0.6                                                                                                               & \checkmark                                                                                               \\
\hline
\end{tabular}
		\end{adjustbox}
	
	\label{table:darpae2}

\end{table}

Figure \ref{fig:all-5}-(c)  shows the graph generated using \projname from the report \cite{deputydog}. The figure demonstrates various system calls executions with specific system entities and the asterisks processes ({\tt *} inside the rectangle). The figure shows several important attacker activities and how they are connected. 
The graph disconnectedness is due to the writing style where the author referred to the same entity by very different names in separate sentences. For instance, the two nodes {\tt *} and {\tt 8aba4b5184072f2a50cbc5ecfe326701} represent the same entity but they are separated in the graph. 

Figure \ref{fig:all-5}-(d), in the Appendix, shows the graph generated using \projname from the report \cite{DustySky}.
Due to the text complexity,  \projname generated three false positive edges {\em 16.exec, 17.exec, and 18.exec}.
Figures \ref{fig:all-5}-(a), \ref{fig:all-5}-(b), and \ref{fig:all-5}-(e), in the Appendix, respectively, demonstrate Carbanak, HawkEye, and Uroburos graphs.

In all the cases, POIROT was able to detect the attacks, even in the presence of false positive edges.

\subsection{Evaluation on the DARPA Transparent Computing Dataset.} \label{subsec:darpa}

In this experiment, we utilized the DARPA Transparent Computing campaign dataset to automatically generate the attack behavior graphs from the natural language description of the attack. During these campaigns, red-teams conducted attacks on infrastructure defended by blue teams. These attacks were carried out on four systems, including one client, one mail server, a web server, and an SSH server over a period of a couple of weeks. The text descriptions of the attacks processed by \projname were written by the red-team members as part of the ground truth release of the exercises. These reports are shorter and more concise than those in the public CTI reports. In addition to textual descriptions, they also contain graph representations of the attacks generated by the red-team members. The graphs generated by \projname were compared with these graphs as ground truth.

\begin{table}
	\caption{Precision, recall, and F1-score of the graphs generated from DARPA reports, calculated by comparing edges in the automated graph against those in the DARPA reports.}
	\centering
\begin{adjustbox}{width=0.7\linewidth}
	\begin{tabular}{|c|c|c|c|} 
		\hline
		\textbf{Scenario}   & \textbf{Precision} & \textbf{Recall} & \textbf{F1-Score}  \\ 
		\hline
		Simple APT   & 1.00               & 1.00            & 1.00               \\ 
		\hline
		Micro APT    & 0.88               & 1.00            & 0.94              \\ 
		\hline
		Drakon APT    & 1.00               & 0.84            & 0.91               \\ 
		\hline
		Gather App  & 1.00               & 0.88            & 0.94               \\ 
		\hline
		HelloWorld  & 1.00               & 1.00            & 1.00               \\ 
		\hline
		GatherApp   & 1.00               & 1.00            & 1.00               \\ 
		\hline
		Webshell    & 0.89               & 0.89            & 0.89               \\ 
		\hline
		Metasploit  & 0.91               & 0.91            & 0.91               \\
		\hline
	\end{tabular}
\end{adjustbox}
	\label{table:DARPAFPFN}
 \vspace{-1mm}
\end{table}

\begin{table*}[]
	\caption{The reports and graphs' characteristics averaged across all reports.}
	\centering
	\small
	\begin{adjustbox}{width=1\textwidth}
		\small
		\begin{tabular}{|c|c|c|c|c|c|c|c|c|c|c|} 
\hline
\multirow{2}{*}{\textbf{Scenario}} & \multirow{2}{*}{\begin{tabular}[c]{@{}c@{}}\textbf{Number of }\\\textbf{ Reports } \end{tabular}} & \multirow{2}{*}{\begin{tabular}[c]{@{}c@{}}\textbf{Smallest}\\\textbf{Report } \end{tabular}} & \multirow{2}{*}{\begin{tabular}[c]{@{}c@{}}\textbf{Largest}\\\textbf{Report } \end{tabular}} & \multirow{2}{*}{\begin{tabular}[c]{@{}c@{}}\textbf{Avg. Number }\\\textbf{ of Sentences } \end{tabular}} & \multirow{2}{*}{\begin{tabular}[c]{@{}c@{}}\textbf{Avg. Sentences}\\\textbf{ After Summarization } \end{tabular}} & \multicolumn{2}{c|}{\textbf{Avg. Attack Behavior } } & \multicolumn{2}{c|}{\textbf{Avg. Removal } } & \multirow{2}{*}{\begin{tabular}[c]{@{}c@{}}\textbf{Avg. MCS }\\\textbf{  Score } \end{tabular}}  \\ 
\cline{7-10}
                                   &                                                                                                   &                                                                                               &                                                                                              &                                                                                                          &                                                                                                                             & \textbf{|V| }  & \textbf{|E| }                     & \textbf{|V| }  & \textbf{|E| }             &                                                                                                            \\ 
\hline
Microsoft                          & 4020                                                                                              & 19                                                                                            & 63                                                                                           & 32.26                                                                                                    & 19.02                                                                                                                       & 18              & 17                                 & 7               & 6                          & 0.91                                                                                                       \\ 
\hline
TrendMicro                         & 11600                                                                                             & 17                                                                                            & 59                                                                                           & 31.93                                                                                                    & 14.22                                                                                                                       & 16              & 15                                 & 5               & 4                          & 0.85                                                                                                       \\
\hline
\end{tabular}
	\end{adjustbox}
	\label{table:largescale}
\vspace{-2mm}
\end{table*}

Table \ref{table:darpae2} describes the results of this experiment. For each attack (named in the first column), it shows the report's size in sentences (in the second column) and the manual graph's size generated by the attackers and by \projname.  The differences between the manual and \projname's graphs are minimal due to the shorter size of the CTI reports and their conciseness. Columns seven and eight represent the results of threat hunting where POIROT \cite{milajerdi2019poirot} was used to detect the EXTRACTOR-generated graph inside the provenance graph generated from the audit records, which includes both attack and benign activities. In all scenarios, our detection score surpassed the detection threshold ($t\approx0.3$), and the attack was detected successfully.

Table \ref{table:DARPAFPFN} shows the performance of \projname and its capability in capturing all relevant attack behaviors on the DARPA reports.  The result shows improvement in the performance of \projname on the DARPA CTI reports compared to the public CTI reports (Table \ref{table:CTIFPFN}). This is due to the simplicity of the DARPA reports, which resulted in generating fewer false positives. 
Similar to the CTI reports, most false-negatives are due to the \projname model not being able to drive the relation from the sentence. 

To further examine the possibility of generating false detection signals, we ran POIROT on a benign dataset of audit logs of the DARPA TC program, and provided in input the (attack) graphs extracted by \projname. The dataset includes 12GB benign audit logs from the different operating systems, including Windows, Linux, FreeBSD. Our threat detection using POIROT raised no false signals. This experiment explicitly shows that \projname graphs are concise enough not to raise false detection signals in benign environments.

\subsection{Large Scale Experiment} \label{subsec:largscale}

To evaluate the scalability of our approach and its accuracy with additional writing styles, we process with \projname a large number of unstructured CTI reports from two major CTI sources, namely Microsoft Security Intelligence \cite{msthreat}
and TrendMicro Threat Encyclopedia \cite{trendmicroenc}.

The main challenge in this kind of evaluation is the absence of ground truth. While in the first two cases there were graphs to compare with, these CTI sources do not provide such graphs. However, the reports themselves point to a way to overcome such a challenge, described next. 

Reports from these sources usually contain several sections including threat summary, technical description, and solution where they describe the overview of the attack, the technical attack details, and the steps required to remove the attack. While the first section provides some general information such as infection rate and risk and severity level about the threat, the second and the third sections provide valuable technical insight about the attack and how to reverse its impacts. Often, the last two sections of these reports are similar but antithetical to each other. In other words, while an {\em attack description} section describes the steps taken to compromise a system,  including files created and executed, processes compromised etc, a {\em solution} section details the steps needed to remove the attack's artifacts, i.e., the same files created, and compromised processes. For instance, the sentence \texttt{Delete} 
\path{<system folder>\sysformat.exe } \texttt{ from }
\path{	HKEY\CURRENT\USER\SOFTWARE\Microsoft\Windows\CurrentVersion\Run
} 
from the solution sections outlined the action required to undo the threat action described as  \texttt{Adds registry value: sysformat
	with data: <system folder>\textbackslash sysformat.exe
	in the registry key:} \path{HKEY\CURRENT\USER\Software\Microsoft\Windows\CurrentVersion\Run} in the {\em attack description} section.  As another example, the sentence \texttt{check for the open connection to 10.13.13.1} from the {\em solution} section maps to the \texttt{connects to command and control sever 10.13.13.1} from the {\em attack description} section. We note, at this point, that the {\em solution} section does not have these characteristics across all the reports. Indeed, it often amounts to instructions on how to download and execute a {\em patch} file, which does all the clean-up and patching work. However, it is relatively easy to automatically distinguish between larger {\em solution sections} that contain detailed clean-up steps, and smaller {\em solution} sections that instruct to run a patch file, and filter out the latter. The reports in \cite{BKDR} and \cite{rbot} are examples such CTI reports, while more examples can be found at \cite{trendmicroenc} and \cite{msthreat}.

To evaluate \projname, in this experiment, for each report that contains both an {\em attack description} and a detailed  {\em solution}, we build the provenance graphs related to each section by omitting the other section from the rest of the report. Then we invert the graph obtained from the {\em solution} and calculate its similarity with the graph obtained from the {\em attack description} section. 
To measure the similarity between the two graphs, we use the Maximum Common Subgraph (MCS) \cite{raymond2002maximum}, a metric that measures the containment of a smaller graph inside a larger graph. (We use this metric for this large scale evaluation as it is considerably simpler than the notion of alignment used in ~\cite{milajerdi2019poirot}).  

\definecolor{dor}{HTML}{006B3C} 
\definecolor{tu}{HTML}{7FFFD4}

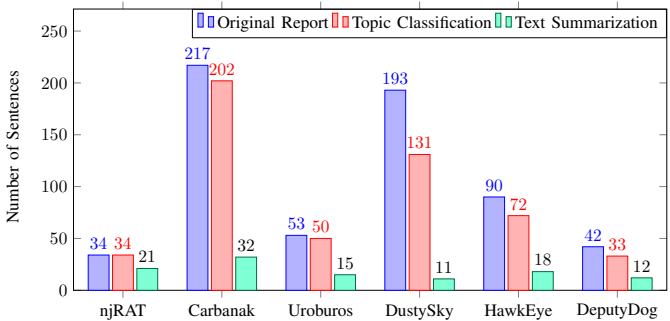
\begin{figure}
	\begin{adjustbox}{width=1\linewidth}
		\begin{tikzpicture}
		\begin{axis}[
		ybar,
		bar width=12pt,
		x=2cm,
		enlarge y limits={0.25,upper},
		ymin=0,
		legend style={at={(1,1)},anchor=north east, legend columns=-1},
		ylabel={\ Number of Sentences},
		symbolic x coords={njRAT, Carbanak, Uroburos, DustySky, HawkEye,DeputyDog},
		xtick=data,
		nodes near coords,
		nodes near coords align={vertical},
		]
		\addplot coordinates {(njRAT,34)(Carbanak,217)(Uroburos,53)(DustySky,193)(HawkEye,90)(DeputyDog,42)};
		\addplot coordinates {(njRAT,34)(Carbanak,202)(Uroburos,50)(DustySky,131)(HawkEye,72)(DeputyDog,33)};
		
		\addplot[draw=dor, fill=tu] coordinates {(njRAT,21)(Carbanak,32)(Uroburos,15)(DustySky,11)(HawkEye,18)(DeputyDog,12)};
		\legend{Original Report,Topic Classification, Text Summarization}
		\end{axis}
		\end{tikzpicture}
	\end{adjustbox}
	\vspace{-2mm}
	\caption{The number of candidate sentences after applying text summarization, compared to the number of sentences in the raw report and the number of sentences after Topic Classification (TC).}
	\label{table:rawtcets}
\end{figure}

Table \ref{table:largescale} shows the results of this experiment. In this table, the second column shows the number of the evaluated reports. The third and fourth columns describe the length of the smallest and largest report. The third and the fourth columns show the average number of sentences before and after text summarization. The average attack description and average solution columns show the average number of nodes and edges build from the technical details and removal section, and finally, the last column measures the similarity between the two graphs. As can be seen, the average similarity measure between the extracted graphs is equal to 0.91 for Microsoft and 0.85 for TrendMicro. This means that \projname correctly extracts the graphs from the text in a majority of the cases.

For further comprehensiveness of this experiment, we also performed manual ‘spot-checks’, where we manually evaluated 50 randomly chosen reports. The individual assessments were then discussed and agreed upon in a meeting. 
The false-negatives and false-positives are generated due to the unresolved complexities and are still minimal, considering the complexity of the report, and are in line with our previous result. Our precision, recall and F1-score was 0.88, 0.93, 0.90.

\subsection{Fine Grained Performance Evaluation} \label{subsec:ETS}

In this subsection, we provide more fine-grained evaluations on some of the most important steps of our approach. 

\noindent\textbf{Text Summarization.} Previous work in automatic extraction of knowledge from CTI reports \cite{liao2016acing,husari2017ttpdrill} uses topic classification (TC) to discern irrelevant content such as advertisements from the CTI reports. In \projname, we use a different approach for text summarization aimed at achieving a finer-grained summarization. Figure \ref{table:rawtcets} shows these two approaches side by side. To do topic classification, similarly to the previous approach \cite{liao2016acing}, we ran a Support Vector Machine (SVM) classifier on 1500 technical CTI sentences versus 1000 sentences of advertisement and about author details. We evaluated the model using 10-fold cross-validation, achieving a precision of 97\% and a recall of 99\%. As can be seen in the figure, our text summarization approach significantly reduces the size of the text compared to Topic Classification.

\begin{table}[!t]
		\caption{The performance evaluation of CNN and LSTM neural networks versus  BERT language model on sentence-verbosity task. The best result are bolded.}
	\centering
	\small
	\begin{adjustbox}{width=0.7\linewidth}
	\begin{tabular}{|c|c|c|c|}
		\hline  
		\textbf{Scenario} & \textbf{Precision }     & \textbf{Recall }        & \textbf{F-1 Score}      \\ \hline
		CNN   & 0.895          & 0.895          & 0.897          \\ \hline
		LSTM  & 0.883          & 0.894          & 0.887          \\ \hline
		BERT  & \textbf{0.950} & \textbf{0.957} & \textbf{0.953} \\ \hline
	\end{tabular}
\end{adjustbox}
	\label{table:ets}
\end{table}

\begin{table}[!t]
	\caption{Each cell represents the MCS against the baseline graph generated by \projname, without (w/o) activating the corresponding module. One denotes results that are the same as the baseline, while zero signifies no match between baseline verses the graph drawn without the corresponding module. 
	} 
	\small
	\centering
\begin{adjustbox}{width=0.9\linewidth}
\begin{tabular}{|c|c|c|c|c|c|} 
		\hline
		\multirow{2}{*}{ \textbf{Scenario} } & \textbf{w/o}                        & \textbf{w/o~}                       & \multicolumn{3}{c|}{\textbf{Resolution} }                \\ 
		\cline{4-6}
		& \multicolumn{1}{l|}{\textbf{Toke.}} & \multicolumn{1}{l|}{\textbf{Homo.}} & \textbf{w/o ESR}  & \textbf{w/o PR}  & \textbf{w/o ER}   \\ 
		\hline
		njRAT                                & 0.40                                & 0.28                                & 0.50              & 0.90             & 0.81              \\ 
		\hline
		Carbanak                            & 1.00                                & 0.42                                & 0.88              & 0.88             & 0.97              \\ 
		\hline
		Uroburos                             & 0.85                                & 0.41                                & 1.00              & 0.85             & 1.00              \\ 
		\hline
		DustySky                             & 0.71                                & 0.0                                 & 1.00              & 1.00             & 1.00              \\ 
		\hline
		HawkEye                              & 0.90                                & 0.15                                & 0.90              & 0.83             & 1.00              \\ 
		\hline
		DeputyDog                            & 1.00                                & 0.16                                & 1.00              & 1.00             & 0.83              \\
		\hline
	\end{tabular}
\end{adjustbox}
\label{table:aptabelation}

\end{table}

Finally, Table \ref{table:ets} presents the result of the {\em Sentence Verbosity} removal using the states of the art approaches, attesting that BERT outperforms other popular models.

\noindent\textbf{Ablation Study.}
To demonstrate the contribution of each \projname module toward the final graph, we performed an ablation study to measure the similarity of the graph generated in the absence of that module compared to the baseline (having all modules active). 

Table \ref{table:aptabelation} shows results of our ablation study. Each column represents the result of the MCS similarity score  of EXTRACTOR's generated graph in the absence of that component(w/o). The baseline for this study is the overall performance of EXTRACTOR, set at 1. Each column shows the loss of performance when any specific component is omitted in the overall approach.  The table also shows the diversity in the writing styles and the fact that every single technique matters (as there is no column with all 1’s), showing the need to combine these different techniques across the various reporting scenarios. Our results show that all the different modules  in \projname successfully contribute to various degrees, depending on the text style, to the concise graph generation. In particular, they enable \projname to process a wide variety of writing styles successfully.

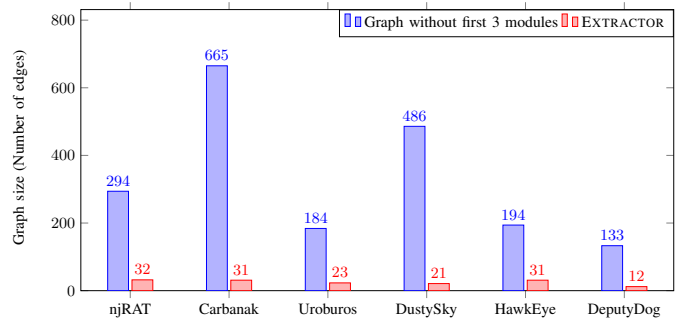
\begin{figure}[t]

	\centering
	\small
	\begin{adjustbox}{width=1\linewidth}
		\begin{tikzpicture}
		\begin{axis}[
		ybar,
		bar width=12pt,
		x=2cm,
		enlarge y limits={0.25,upper},
		ymin=0,
		legend style={at={(1,1)},anchor=north east, legend columns=-1},
		ylabel={\ Graph size (Number of edges)},
		symbolic x coords={njRAT, Carbanak, Uroburos, DustySky, HawkEye,DeputyDog},
		xtick=data,
		nodes near coords,
		nodes near coords align={vertical},
		]
		
		\addplot coordinates {(njRAT,294)(Carbanak,665)(Uroburos,184)(DustySky,486)(HawkEye,194)(DeputyDog,133)};
		\addplot coordinates {(njRAT,32)(Carbanak,31)(Uroburos,23)(DustySky,21)(HawkEye,31)(DeputyDog,12)};
		
		\legend{Graph without first 3 modules ,\projname}
		\end{axis}
		\end{tikzpicture}
	\end{adjustbox}

	\caption{The size of the generated graph before and after utilizing Normalization, Resolution, and Summarization modules.}
	\label{fig:comparision}
\end{figure}

In addition, we examined the impact of the steps of Normalization, Resolution, and Summarization, by not performing these steps and rather building the graphs from the raw reports. Figure \ref{fig:comparision} shows that the size of such graphs is in the hundreds of edges, while the size of the graphs obtained by the full chain of modules is much smaller. 

Table \ref{table:prevekance} shows the prevalence of challenges discussed in Section \ref{sec:background} in two of the major threat report websites \cite{msthreat} and \cite{trendmicroenc}. Each column represents the number of times that each module has been invoked (we avoid adding tokenizer as it has been reflected in Table \ref{table:largescale}). The second column shows the total number of analyzed reports. The third and the fourth columns outline the average number of homogenized instances and passive to  active conversation, respectively. Finally, the resolution column presents the results of ESR, PR, and ER. 

Table \ref{table:basevsretrained} presents the SRL's performance before and after retraining. Finally, Table \ref{table:SEE} shows the performance of the SEE module in picking correct and meaningful nodes.
To evaluate our SEE's completeness, we ran our SEE module on 1,000 public reports used by \cite{zhu2018chainsmith} and compared SSE results against their result as a baseline.  Table \ref{table:SEE} presents the result of this evaluation.

\definecolor{dor}{HTML}{006B3C} 
\definecolor{tu}{HTML}{7FFFD4}

\begin{table}[!t]
	\caption{The average number of times each module invoked. Table shows the prevalence of challenges discussed in Section \ref{sec:background}. {\em Homo} refers to {\em Homogenization}. }
	\small
	\centering
	\begin{adjustbox}{width=0.9\linewidth}
	\begin{tabular}{|c|c|c|c|c|c|c|} 
		\hline
		\multirow{2}{*}{ \textbf{Scenario} } & \multirow{2}{*}{\begin{tabular}[c]{@{}c@{}}\textbf{\# of}\\\textbf{Reports} \end{tabular}} & \multirow{2}{*}{\begin{tabular}[c]{@{}c@{}}\textbf{Homo.}\end{tabular}} & \multirow{2}{*}{\begin{tabular}[c]{@{}c@{}}\textbf{Passive to}\\\textbf{Active}\end{tabular}} & \multicolumn{3}{c|}{\textbf{Resolution} }   \\ 
		\cline{5-7}
		&                                                                                                &                                                                                                      &                                                                                               & \textbf{ESR}~ & \textbf{PR}~ & \textbf{ER}  \\ 
		\hline
		Microsoft                            & 4020                                                                                           &             27.90                                                                                         &      2.07                                                                                       &       5.49        &      4.59        &   4.60           \\ 
		\hline
		TrendMicro                           & 11600                    &     23.32                                  &            1.34                                                                                  &        2.52     &       5.69       &   5.31           \\
		\hline
	\end{tabular}
\end{adjustbox}
	\label{table:prevekance}
\end{table}

\begin{table}[!t]
	\caption{The performance evaluation before and after training the SRL model on  CTI data, where we used 80\%, 20\%, 20\% for the train, test, and validation, respectively.}
	\centering
	\small
	\begin{adjustbox}{width=0.67\linewidth}
		\begin{tabular}{|c|c|c|c|}
			\hline  
			\textbf{Scenario} & \textbf{Precision }     & \textbf{Recall }        & \textbf{F-1 Score}      \\ \hline
			SRL(Base)   & 0.83         & 0.82          & 0.84          \\ \hline
			SRL(Retrained)  & 0.92 &0.93 & 0.93 \\ \hline
		\end{tabular}
	\end{adjustbox}
	\label{table:basevsretrained}
\end{table}

\begin{table}[!t]
	\caption{The performance of SEE module.}\vspace{-2mm}
	\centering
	\small
	\begin{adjustbox}{width=0.57\linewidth}

		\begin{tabular}{|c|c|c|c|}
			\hline  
			\textbf{Scenario} & \textbf{Precision }     & \textbf{Recall }        & \textbf{F-1 Score}      \\ \hline
            SEE  & 1         & 0.98          & 0.99          \\ \hline
		\end{tabular}
	\end{adjustbox}

	\label{table:SEE}
\end{table}

\section{Discussion and Limitations}\label{sec:discussion}

\noindent\textbf{False Positives and False Negatives.} 
As  shown in Tables \ref{table:CTIFPFN} and \ref{table:DARPAFPFN} there are extraneous nodes or edges in the \projname generated graphs. As discussed in Sections \ref{sec:background} and \ref{sec:methodology},
we note that some loss of precision in extraction is inevitable due to general issues in dealing with natural language complexity. What is, therefore,  the significance of these extraneous nodes and edges (or missing ones) in the \projname generated graph with respect to the `big picture' of the threat hunting problem? To answer this, we note our choice of a threat hunting approach that uses approximate matching~\cite{milajerdi2019poirot}, facilitates us to successfully identify the threat despite the extraneous information. In fact, we can go on to argue that such approximate techniques are required of the general threat-hunting problem, as it is likely that not all of the activities described in a threat report are likely to manifest in a host due to intrinsic factors (e.g., the non-deterministic factors such as randomness or time affecting the execution of a threat binary) in the activity-based provenance graphs. Therefore, the approximation that is inherently needed for threat-hunting is able to work with the small loss of precision inherent to NLP, as shown in Section \ref{sec:evaluation}. 

\noindent\textbf{Limitations and Future Works.} 
 \projname' s performance may diminish in scenarios where the description of an action spans several sentences or a paragraph, where subject or object resolution might face challenges based on how the threat report was authored. 
 As an instance of this challenge, see the discussion about Figure \ref{fig:all-5}-(c) in Section \ref{subsec:publicctireport}. \projname may deal with this issue if additional  information in the form of alternate reports is made available to resolve these challenges. Another way to resolve this is to facilitate mechanisms  in \projname to actively collaborate with the human operator  to resolve these entities. 

Another limitation of our graph-based approach is that it is not applicable to attacks that involve timing, side-channel inference, etc. The graphs describing the attack behavior are modelled after audit logs that do not capture information at a granularity that enables these attacks to be detected.  However, this limitation is common with other approaches that involve provenance graph-based detection or threat hunting. Similarly, our approach only extracts the attack behaviors described in the natural language and cannot resolve the behaviors represented in other forms like figures and charts.   

Various modules of \projname use dictionaries to enhance the overall performance. While we have tried to be complete in choosing appropriate words, there may be reports where people use words that are not in the dictionaries. Therefore, there can be room to improve the dictionaries further. Using the Named Entity Recognition (NER) system may also enhance the approach in translating noun phrases into unified system representations. 
Moreover, future works may extend the EXTRACTOR to infer the graph from unstructured vulnerability reports. These graphs can be further used to detect possible vulnerabilities within the application. 

\section{Related Work}\label{realated}

\noindent\textbf{Provenance  Graph Analytics. }
Several research projects utilize system audit logs for attack reconstruction and forensic analysis, and threat hunting \cite{hossaincombating, goel2005forensix,goel2005taser,milajerdi2019poirot,liu2018towards,pohly2012hi}. 
Hercule \cite{pei2016hercule} rebuilds attack stages through comparing logs collected from various sources. 
Bilge et al. \cite{bilge2012disclosure} leverage NetFlow logs for detecting botnet C\&C channels. Oprea et al. \cite{oprea2015detection}  uses web proxy and DNS logs to identify infections in enterprise networks.

King et al. \cite{king2005enriching,king2003backtracking} introduced the practice of constructing provenance graphs from kernel audit logs.
Several studies have used provenance graphs in attack detection and forensics \cite{hossain2018dependence, lee2013loggc, xu2016high, hassan2018towards}. 
Hossain et al. \cite{hossain2018dependence},  Xu et al.  \cite{xu2016high}, and LogGC \cite{lee2013loggc} proposed reduction techniques that reduce the size of the graph while maintaining the accuracy needed for forensic analysis. \projname can be a companion to these approaches to provide a clear picture of attacks.

\noindent\textbf{NLP and Threat Information Extraction.} 
Several open standards such as STIX \cite{stix}, MISP \cite{platformopen}, and OpenIOC \cite{openioc} have been proposed to exchange knowledge about IOCs in an interoperable way. However, unlike our approach, these standards are more focused on exchanging IOCs than describing how those IOCs are connected and how the attacks behave (see the examples at \cite{stixexample}). 
	Companies that use threat exchange standards instead of only relying on the exchange of high-level threat data can benefit from the publicly available knowledge in the wild. On the other hand, these threat exchange standards are limited in usage, as companies are not equally interested in sharing their data. Moreover, the exchanged data does not contain technical details such as the affected registry, file path, and the application name as they can be a privacy leakage of the company's private information.  As in many cases, the organizations' internal policies prevent the sharing of data with outside entities \cite{sadique2018automated,de2017pracis}.

The VirusTotal Graph \cite{virusgraph} also differs from our work as it only represents the high-level view of the attack, mainly including hashes, IPs, and domains involved in a possible threat scenario. Also, unlike our approach, VirusTotal's report is generated based on the analysis of sample malware, while \projname by having access to the publicly available reports (which can include the VirusTotal) allows utilization of public CTI, converting raw reports into actionable knowledge. 

iACE \cite{liao2016acing} proposes a graph mining approach to extract IOCs from security articles. ChainSmith \cite{zhu2018chainsmith} uses NLP to extract IOCs from security articles and further categorize them into campaign stages.  
TTPDrill \cite{husari2017ttpdrill} proposes an ontology which helps to understand the characteristics and specifications of cyber threats. It uses NLP and Information retrieval (IR) to extract the threat actions from reports. The work of \cite{husari2019learning} creates TTP chains from reports, using DP rules.  
Unlike these approaches, \projname focuses on extracting the attack behavior and captures system-level causality in the form of the provenance graphs. 

SemFuzz \cite{you2017semfuzz} performs fuzzing guided by  information extraced from vulnerability reports. 
Feng et al. \cite{feng2019understanding} use NLP to generate network signatures from unstructured vulnerability reports. They use those signatures in intrusion detection and firewall systems.
Dong et al. \cite{dong2019towards} use Named Entity Recognition and Relation Extraction  to extract software name and version and report inconsistency between major vulnerability databases. 
Even though, somehow related, \projname's goal and techniques are essentially different from these works.

Featuresmith \cite{zhu2016featuresmith}  generates a feature set for detecting Android malware from security literature. In contrast, \projname aims to build a provenance graph that represents the actual behavior of the attack. 
Privee \cite{zimmeck2014privee} leverages machine learning to retrieve web policies. The work of \cite{qu2014autocog} and \cite{pandita2013whyper} relate the app description with permissions using NLP. The works of \cite{nan2015uipicker} and \cite{huang2015supor} identify users' sensitive inputs in Android app. EKLAVYA \cite{chua2017neural} uses NLP to recover function signatures from binary code.  

\section{Conclusion}\label{conclusion}

\projname automatically builds a provenance graph from CTI reports written in natural language. 
  We evaluate \projname using various threat reports and real-world attack scenarios. \projname successfully extracts graphs that match those drawn manually by security experts, and those graphs were successfully used for threat detection. 

\section{acknowledgments}

\noindent We would like to thank reviewers for their helpful review comments and suggestions to the manuscript. Thanks are especially due to our anonymous shepherd who provided many useful suggestions  for refinement. This work was supported by DARPA under SPAWAR (N6600118C4035) and NSF (CNS-1918542). The views, opinions, and/or findings expressed are those of the authors and should not be interpreted as representing the official views or policies of the U.S. Government.

\bibliographystyle{IEEEtrans}
\bibliography{bibliography}

\section{Appendix} \label{sec:appendix}

\begin{table}[!h]
	\caption{Sample of noun phrases and their corresponding synonyms in noun dictionary.}
	\centering
	\small
		\begin{adjustbox}{width=0.80\linewidth}
	\begin{tabular}{|c|c|} 
		\hline
		\textbf{Noun}  & \textbf{Synonyms}                                                                                                                                                                                 \\ 
		\hline
		\textbf{IP:.*}           & \begin{tabular}[c]{@{}c@{}}'CC server', 'CC', 'command and control sever',\\ 'C2 server', 'c2', 'candc server', 'C2 node', 'CandC', \\ 'CandC', 'command controle sever', 'C2', 'CandC server',\\  'CC server', 'CommandControle sever', 'Command Controle',... \end{tabular}        \\ 
		\hline
		\textbf{TEMP}             & \begin{tabular}[c]{@{}c@{}}'\%TEMP\%,  '<TEMP> ', 'Windows temporary folder',\\ 'temporary folder' , '\%Temporary folder\%, 'TMP',\\ '\%Temp Folder\%' '\%Temp directory\%,... \end{tabular}                                                                                                                     \\ 
		\hline
		\textbf{Home Folder}      & \begin{tabular}[c]{@{}c@{}}'\%HOMEPATH\%', '\textless{}HOMEPATH\textgreater{}', \\ '\%HOME\_PATH\%', '\textless{}HOME\_PATH\textgreater{}', '\%HOME\%', \\'\textless{}HOME folder\textgreater{}', \textless{}HOME Directory\textgreater{}',\\ 'USER PATH', '\%USER Directory\%',...\end{tabular}  \\
		\hline
	\end{tabular}
	\end{adjustbox}
\label{noun}
\end{table}

\begin{table*}[!h]
\caption{Malware reports details and characteristics.}
\centering
\begin{adjustbox}{width=1\linewidth}
\begin{tabular}{|c|c|c|c|c|c|c|} 
\Xhline{2\arrayrulewidth}
 \textbf{Scenario}  & \textbf{Year} & \textbf{Risk} & \begin{tabular}[c]{@{}c@{}}\textbf{Submitted }\\\textbf{Samples}\end{tabular} & \textbf{Primary target}                                                                   & \textbf{Malware MD5}             &  \textbf{Description}~~                                                                                                                                                                                                                                             \\ 
\hline
njRAT \cite{solutions2013njrat}  - fig. \ref{fig:njrat}          & 2013     & High      & 30                                                                            & \begin{tabular}[c]{@{}c@{}}Middle eastern governments,\\energy sectors, and telecom\\industries.\end{tabular}     & 2013385034e5c8dfbbe47958fd821ca0          & \begin{tabular}[c]{@{}c@{}}The malware has several capabilities, including logging keystrokes, uploading and downloading files, \\recording the victim's camera, steal user credentials stored in the system, open a reverse shell, and\\~manipulations~of the process, file, and registry, etc.\end{tabular}  \\ 
\hline
Carbanak  \cite{carabanak}   - fig. \ref{fig:all-5}-(a)     & 2015   & High        & 109                                                                           & \begin{tabular}[c]{@{}c@{}}Banking and financial \\institutions\end{tabular}                                                        & 1e47e12d11580e935878b0ed78d2294f           & \begin{tabular}[c]{@{}c@{}}The malware has several capabilities, including logging keystrokes, uploading and downloading files, \\recording the victim's camera, steal user credentials stored in the system, open a reverse shell, and\\~manipulations~of the process, file, and registry, etc.\end{tabular}  \\ 
\hline

Uroburos    \cite{uroburos} - fig. \ref{fig:all-5}-(e)      & 2014  & High         & 4                                                                             & \begin{tabular}[c]{@{}c@{}}"the most significant breach of\\ U.S. military computers"\cite{uroburos}\end{tabular} & 51e7e58a1e654b6e586fe36e10c67a73           & \begin{tabular}[c]{@{}c@{}}The malware exploits vulnerabilities in Java (CVE-2012-1723), Adobe Flash (unknown) \\or Internet Explorer 6, 7, 8 exploits (unknown), and is capable~of performing a wide range of tasks.\end{tabular}                                                                             \\ 
\hline

DustySky     \cite{DustySky}  - fig. \ref{fig:all-5}-(d)   & 2015  & High         & 79                                                                            & \begin{tabular}[c]{@{}c@{}}Intelligence gathering \\with Political motives\end{tabular}                                              & 0756357497c2cd7f41ed6a6d4403b395           & \begin{tabular}[c]{@{}c@{}}The malware is written in .NET by a politically-motivated group with primary targets in the \\Middle East, Europe, and the United States and can collect a wide range of details from the target system.\end{tabular}                                                               \\ 
\hline 

HawkEye    \cite{fortinet}  -  fig. \ref{fig:all-5}-(b)      & 2019 & High          & 3                                                                             & \begin{tabular}[c]{@{}c@{}}~A wide range of industries \\and sectors\end{tabular}                                                    & 666a200148559e4a83fabb7a1bf655ac           & \begin{tabular}[c]{@{}c@{}}The malware has several capabilities, including stealing email credentials, logging keystrokes, \\taking screenshots, USB propagation, stealing Bitcoin wallet info, Antivirus, firewall checking, etc.~\end{tabular}                                                               \\ 
\hline

DeputyDog   \cite{deputydog}   - fig. \ref{fig:all-5}-(c)    & 2013   & High        & 8                                                                             & Against Japanese Targets                              & 8aba4b5184072f2a50cbc5ecfe326701           & ~ZeroDay~CVE-2013-3893 against Microsoft~internet explorer - Japan                                                                                                                                                                                                                                             \\
\Xhline{2\arrayrulewidth}
\end{tabular}
\end{adjustbox}
\label{table:malwares}
\end{table*}

\begin{figure*} []
	\centering
	\includegraphics[width=1\textwidth]{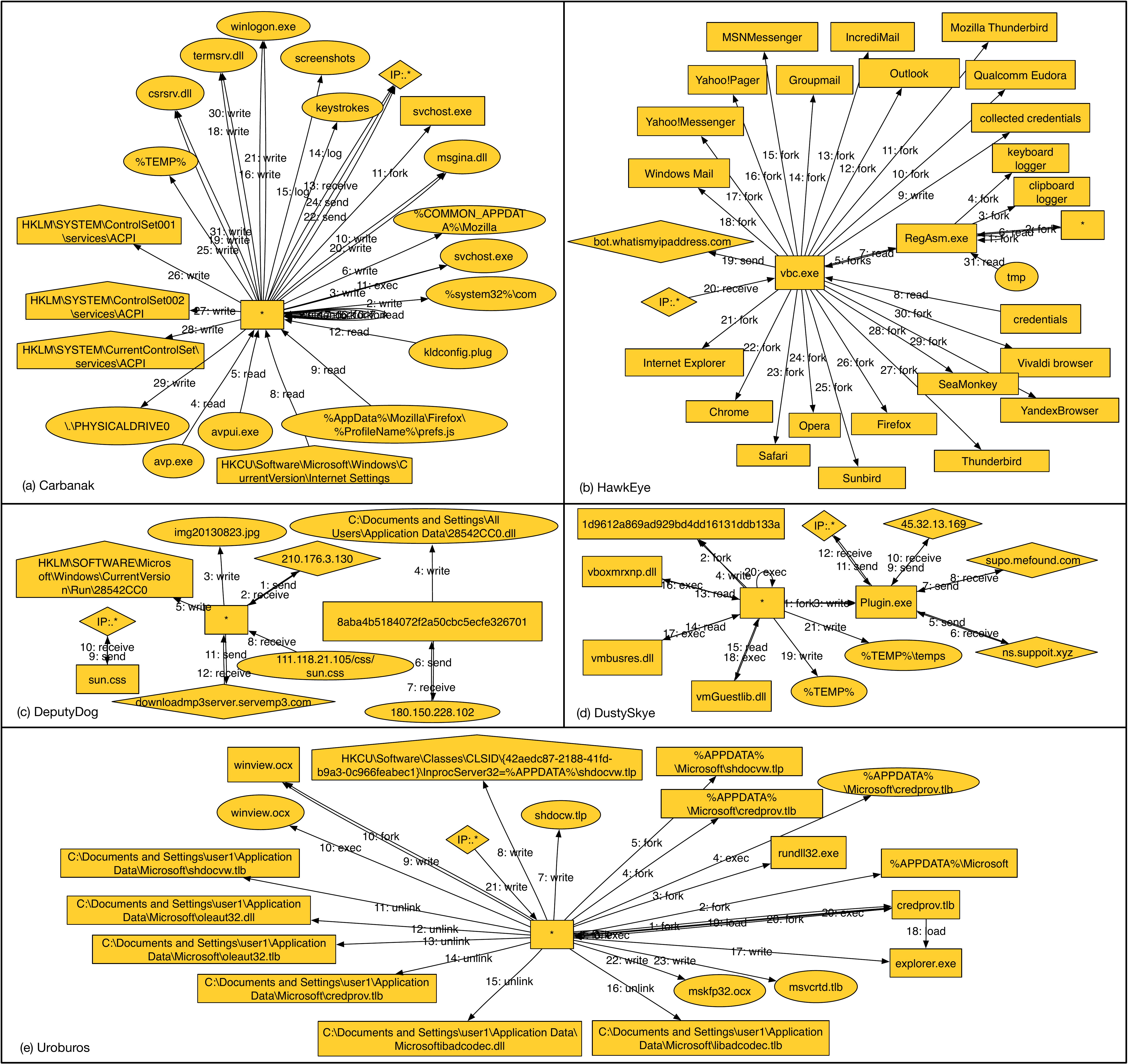}
	\vspace{-4mm}
	\caption{Graphs after generalization, keeping IOCs and asterisking unknown system entities. }
	\label{fig:all-5}
\end{figure*}
	
\end{document}